\documentclass[preprint,aps,prd,preprintnumbers,nofootinbib]{revtex4}
\usepackage{amssymb,amsmath,mathbbol} 

\def\hat{\widehat}
\def\tr{\textrm{tr}}
\def\Tr{\textrm{Tr}}

\def\half{{\textstyle{1\over2}}}

\def\CA{{\cal A}}
\def\CB{{\cal B}}
\def\CC{{\cal C}}
\def\CD{{\cal D}}

\def\CG{{\cal G}}
\def\CH{{\cal H}}

\def\CN{{\cal N}}
\def\CO{{\cal O}}
\def\CP{{\cal P}}
\def\CQ{{\cal Q}}
\def\CS{{\cal S}}

\def\CV{{\cal V}}

\def\spose#1{\hbox to 0pt{#1\hss}}
\def\ltapprox{\mathrel{\spose{\lower 3pt\hbox{$\mathchar"218$}}
 \raise 2.0pt\hbox{$\mathchar"13C$}}}
\def\gtapprox{\mathrel{\spose{\lower 3pt\hbox{$\mathchar"218$}}
 \raise 2.0pt\hbox{$\mathchar"13E$}}}
\def\inapprox{\mathrel{\spose{\lower 3pt\hbox{$\mathchar"218$}}
 \raise 2.0pt\hbox{$\mathchar"232$}}}

\preprint{LA-UR 05-8131}
\begin{document}

\title{Improved bilinears in lattice QCD with 
non-degenerate quarks}

\author{Tanmoy Bhattacharya and Rajan Gupta}
\affiliation{
Los Alamos National Lab, MS B-285, Los Alamos,
                New Mexico 87545, USA}
\author{Weonjong Lee}
\affiliation{
  School of Physics,
  Seoul National University,
  Seoul, 151-747, Republic of Korea
  }
\author{Stephen R. Sharpe and Jackson~M.~S.~Wu\footnote{Present address:
TRIUMF, 
4004 Westbrook Mall,
Vancouver, BC, V6T 2A3, Canada}}
\affiliation{%
Physics Department, University of Washington,
Seattle, WA 98195-1560}

\date{\today}

\begin{abstract}
We describe the extension of the improvement program for bilinear
operators composed of Wilson fermions to non-degenerate dynamical quarks.
We consider two, three and four flavors, and
both flavor non-singlet and singlet operators.
We find that there are many more improvement coefficients than with degenerate
quarks, but that, for three or four flavors, nearly all can be determined
by enforcing vector and axial Ward identities. The situation is
worse for two flavors, where many more coefficients remain undetermined.
\end{abstract}


\maketitle

\section{Introduction}
\label{sec:intro}

Simulations of lattice QCD with light dynamical quarks are
greatly facilitated by the use of improved actions and operators.
Calculations are underway using various types of
improved fermions---staggered, Domain Wall/overlap, 
maximally twisted and improved Wilson fermions.
Here we focus on improved Wilson fermions, and investigate
how the improvement program can be extended to remove errors
proportional to $a m_q$ ($a$ is the lattice spacing, and $m_q$
a generic quark mass) in the realistic case of non-degenerate
quark masses. Of particular interest are the ``$N_f=2+1$'' theories
with $m_d=m_u<m_s$ and ``$N_f=2+1+1$'' theories with
$m_d=m_u<m_s<m_c$, and we consider both theories here.
As we review below, the improvement of the action
for such theories including $a m_q$ terms has already been considered, 
but the improvement of operators has not. 
As a first step, we consider here the improvement of all quark bilinears.
These are of considerable phenomenological interest,
since their hadronic matrix elements, 
combined with experimental results for form factors,
determine elements of the CKM matrix.
Errors in such hadronic matrix elements proportional to $am_s$ and
particularly $a m_c$ can be significant, and it is thus
important to reduce or remove them.

At present, simulations with staggered fermions are able to reach
the smallest dynamical quark masses. This comes at the cost,
however, of a  multiplication of fermion species, and the
concomitant need to use the fourth root of the fermion determinant,
so that unitarity can at best be restored in the continuum limit.
Wilson fermions have the advantage of a straightforward relation
to the continuum theory: each
lattice fermion gives rise to a single continuum flavor.
They also come, however, with disadvantages:
in their original form, the leading discretization errors are of $O(a)$, 
as compared to $O(a^2)$ for staggered fermions,
and these discretization errors explicitly
break chiral symmetry.
As explained in seminal papers by the ALPHA collaboration~\cite{alpha},
one can apply the Symanzik improvement program to Wilson fermions
and systematically reduce the errors from $O(a)$ to $O(a^2)$.\footnote{%
Since the gauge action gives rise to errors of $O(a^2)$ we do not
consider its improvement here, although our considerations apply
equally well for an improved gauge action, as explained in appropriate
places in the following.}
We recall that this requires the addition
of all dimension five operators to
the action that are consistent with the symmetries of the
lattice theory, with their coefficients being determined
by appropriate non-perturbative conditions.
These conditions are generically chosen to enforce a symmetry
that is present in the continuum limit but is broken for 
non-vanishing lattice spacing.
For Wilson fermions, the broken symmetries that are used are
the flavor non-singlet axial symmetries.

A similar method holds for the improvement of operators.\footnote{%
We use the term ``improvement'' as shorthand for 
``$O(a)$ improvement'' throughout this paper. 
We do not consider the removal of discretization
errors proportional to $a^2$ or higher powers.}
One must add all operators with the same symmetries having
one higher dimension and determine their coefficients by applying
appropriate ``improvement conditions''.
This assumes that the operators do not mix with other operators
of lower dimension, which will be true in all but one case here.

It is possible to simplify this procedure by considering only
the improvement of on-shell quantities---masses, decay constants,
physical matrix elements, etc. This allows one to use the
equations of motion to reduce the number of higher dimension
operators that need be considered.
On-shell improvement is equivalent to improving
correlation functions in which the arguments are all separated.
Off-shell improvement extends this to correlation functions in which
some arguments are at the same space-time point,
and requires additional contact terms. 
While we will consider only on-shell improvement here, it will turn
out that we need to understand some of the subtleties of
off-shell improvement in order to resolve certain puzzles that 
emerge from our analysis.

The general discussion of Refs.~\cite{alpha} applies
in the presence of two or more dynamical quarks. 
Thus we know from that work
how to non-perturbatively improve the action, and it is
now standard to implement this in unquenched simulations.
Results are available with the  Wilson gauge action for 
two~\cite{JansenSommer,Yamada:2004ja}
three~\cite{Yamada:2004ja} and four~\cite{Yamada:2004ja} flavors,
and also with an improved gauge action~\cite{Aoki:2005et}.
The method of Ref.~\cite{alpha} also allows one to improve
the flavor non-singlet axial current in the chiral limit
(and a variant of this method
has been applied for two flavors in Ref.~\cite{DellaMorte:2005se}),
and the methodology has been extended to the improvement of
non-singlet vector and tensor bilinears 
in the chiral limit~\cite{GuagnelliSommer,WIPLB,Martinelli}.
Theoretical discussion of improvement of non-singlet bilinears away 
from the chiral limit has been restricted to 
degenerate quarks~\cite{alpha,Martinelli},
or to non-degenerate quarks in the quenched 
approximation~\cite{Petronzio,WIPLB}.

Here we generalize previous work by considering the
improvement of bilinears for the realistic case
of non-degenerate quarks. We consider both
flavor singlet and non-singlet bilinears; the addition of flavor
singlets is required
by the analysis, but is also of phenomenological interest.
We explain how working away from the chiral limit
introduces a plethora of new improvement constants, and then
study which of these can be determined by imposing appropriate
axial Ward identities. We find that, for three flavors,
nearly all can be determined, with the situation unchanged for
four  flavors but worse for two flavors.
To determine the remaining improvement coefficients 
one must use other methods, e.g. 
non-perturbative renormalization~\cite{NPR},
suitably improved~\cite{CNGI}, 
or matching short distance correlation functions
to perturbative expressions~\cite{Martinelli}.

In this work we need make no specific choice as to how the Ward identities
are implemented. One could use a method based on the
Schr\"odinger functional, 
or use standard hadron correlation functions. 
Our theoretical discussion holds equally well for either choice.
What is key, however, is that one can vary the quark masses independently
in a regime where all have $m_q a < 1$, so that
effects proportional to $(m_q a)^2$ can be neglected.
This condition is satisfied for the 
physical strange quark for lattice spacings satisfying
approximately $a \le (1\ {\rm GeV})^{-1}$,
and for the physical charm quark if $a \le (4\ {\rm GeV})^{-1}$.
We stress that these conditions {\em do not necessarily require the quarks to be
light} (with ``light'' meaning $m_q\ll\Lambda_{\rm QCD}$).

The practical implementation of our method will be very involved and
computationally expensive. An indication of this is
that even the simplest step
of determining the improvement constant for the
non-singlet axial current
in the chiral limit has only recently
been undertaken for dynamical fermions~\cite{DellaMorte:2005se}.
For most improvement constants, present
 calculations instead rely upon one-loop perturbative values.
This raises the question of whether our analysis is of purely theoretical
interest or will be useful in practice. 
To ask this another way, will a tree-level or one-loop perturbative
estimate of the improvement constants suffice in practice?
For example, the mass dependent improvement coefficient for the
non-singlet flavor off-diagonal axial current 
(needed to determine decay constants)
enters in an overall factor of $1 + a b_A (m_j + m_k)/2$.
If we use the tree level value of $b_A=1$, then the error
in this factor is $\sim \alpha_s a m_s/2$, assuming
a one loop correction to $b_A$ of order unity times $\alpha_s$.
Taking $m_s\approx 0.1\;$GeV, $a^{-1}\approx 2\;$GeV, and
$\alpha_s\approx 0.3$, the error is less than 1\%.
This may be smaller than other sources of error, in which case the
tree level value for $b_A$ would suffice.
On the other hand, if we consider $f_D$ in the four flavor theory,
then the corresponding error is much larger, $\sim 10\%$,
and a more accurate determination of $b_A$ is likely needed.

The outline of this paper is as follows. In the next section we recall
previous work on the improvement of the unquenched theory in the chiral 
limit. In section~\ref{sec:coeffs} we describe the additional
improvement coefficients that are needed in the unquenched theory 
for non-vanishing quark masses. We then, in section~\ref{sec:WI}, lay out the
Ward identities that can be used to determine most of these coefficients.
We end by discussing some implications of our results in section~\ref{sec:conc}.
Two appendices present the generalizations to two and four flavors.

This paper is an expansion, clarification and, to some extent, a correction
of Ref.~\cite{lat99}. In that work we argued that some of the
improvement coefficients could be determined by imposing vector Ward
identities. It turns out that this was not correct in all cases, due
to the presence of certain contact terms. We explain this point in
a final appendix. 

A brief summary of the present work has been given in Ref.~\cite{lat03}

\section{Review of previous work}
\label{sec:review}

We begin by reviewing previous work on non-perturbative $O(a)$
improvement of unquenched QCD. The ALPHA collaboration~\cite{alpha}
has shown how on-shell improvement of the action can be accomplished
by adding the Sheikholeslami-Wohlert or ``clover'' term, with
appropriately chosen coefficient $c_{SW}$.  Their method
requires a flavor non-singlet axial current,
and thus works if the number of light quarks, $N_f$, is two or greater.
Of course, the resulting value of $c_{SW}$ depends on $N_f$.

We briefly review the method of Ref.~\cite{alpha}, both for
completeness and to introduce our notation.
One considers matrix elements of the improved axial current 
and pseudoscalar density, which, in the chiral limit, take the form
\begin{eqnarray}
\widehat A_\mu^{(jk)} &=& Z_A A_\mu^{(jk),I}
\qquad
A_\mu^{(jk),I} = A_\mu^{(jk)} + a c_A \partial_\mu P^{(jk)} 
\qquad (j\ne k)
\\
\widehat P^{(jk)} &=& Z_P P^{(jk),I}
\qquad
P^{(jk),I} = P^{(jk)}
\qquad (j\ne k)
\,.
\end{eqnarray}
Here we introduce the notation that we use throughout this article: a
``hat'' on an operator indicates that it is both $O(a)$ improved {\em
and} properly normalized, while the superscript ``I'' indicates improvement
alone.  The improvement here is on-shell, not off-shell.  Flavor
indices are shown as superscripts, with $j,k=1-N_f$.
$A_\mu$ and $P$ are ultra-local lattice transcriptions of the axial
current and pseudoscalar density respectively. The simplest choices
are exemplified by
$A_\mu^{(jk)}(x)=\bar\psi^j(x) \gamma_\mu\gamma_5 \psi^k(x)$,
with $a^{3/2}\psi(x)$ being the bare lattice fermion at site $x$,
but our considerations hold for any ultra-local choices.
Finally,
$\partial_\mu$ is an $O(a)$ improved lattice derivative, e.g. the
symmetric difference divided by $2a$. 
Factors of the lattice spacing $a$ are shown
explicitly throughout, so that all quantities have the same dimensions
as their continuum counterparts.

To determine $c_{SW}$ one enforces the simplest axial Ward identity
\begin{equation}
\langle \partial_\mu \widehat A_\mu^{(jk)}(x) \rangle_J =
(\hat m_j + \hat m_k) \langle \hat P^{(jk)}(x) \rangle_J
+ O(a^2) \,.
\label{eq:2ptNSaxial}
\end{equation}
Here the $\hat m_i$ are improved and normalized 
quark masses, whose relation to the
bare quark masses is discussed below.  The subscript on the
expectation values indicates that these matrix elements are to be
evaluated in the presence of sources, $J$, which are located at 
different positions from the operators. The sources should have
quantum numbers such that the result is non-vanishing, but are
otherwise arbitrary (in both form and position).
They could be boundary sources in the Schr\"odinger functional,
or standard hadron operators in a traditional large volume calculation.
We need not (and do not) specify them. 
What matters here is that this equation should hold for any
such sources (which thus create states with the appropriate quantum
numbers in all possible linear combinations). 
The left- and right-hand
sides will only match as the sources are changed
(while holding the bare quark masses and couplings fixed so
that the $\hat{m}_i$ are fixed)
if both $c_{SW}$ and $c_{A}$ are chosen correctly.\footnote{%
In practice, one can cancel the contributions proportional
to one or other of these constants by taking
appropriate linear combinations using different sources.
This has been done in the determination of $c_{SW}$ 
in Refs.~\cite{JansenSommer,Yamada:2004ja,Aoki:2005et},
and of $c_A$ in Ref.~\cite{DellaMorte:2005se}. This allows
one to tune the sources to improve the sensitivity separately
for each improvement constant.}
Since the accuracy of matching is $O(a^2)$, these constants
can only be determined to a relative accuracy of $O(a)$.

An important point concerning the implementation of
eq.~(\ref{eq:2ptNSaxial}) is that one does not need to know $Z_A$,
$Z_P$ or the $\hat m_i$. One need only require that the ratio of the
matrix elements of the improved, but not normalized, quantities,
$\langle A_\mu^{(jk),I} \rangle_J$ and $\langle P^{(jk)} \rangle_J$,
is the same for any choice of $J$. 
One can then
extrapolate the resulting values of $c_{SW}$ and $c_A$ to chiral
limit. This limit can be determined by setting all $N_f$ bare quark masses
equal, and extrapolating to the common value
for which the right-hand side of eq.~(\ref{eq:2ptNSaxial})
vanishes.  
One then has in hand the
desired improvement constants, which will depend,
for a given choice of gauge action, only upon the bare
coupling constant $g_0^2$.\footnote{%
In practice, it may be advantageous not to extrapolate $c_{SW}$
and $c_A$ to the chiral limit~\cite{DellaMorte:2005rd}. There would then be
residual $O(am)$ contributions in these coefficients,
but this does not effect $O(a)$ improvement. We find it
conceptually simpler, however, to imagine that these coefficients
have been extrapolated to the chiral limit, so that they are 
independent of the quark masses that we will be varying in the
subsequent discussion.}

At the same time, one has determined the critical value of hopping
parameter, $\kappa_c(g_0^2)$, as this is the value for which
the right-hand side of eq.~(\ref{eq:2ptNSaxial}) vanishes given degenerate
quarks. 
One can then define bare quark masses in the standard way:
\begin{equation}
a m_j = \frac{1}{2 \kappa_j} - \frac{1}{2\kappa_c}
\,.
\end{equation}
These calculations also give
information about $Z_A/Z_P$, but we postpone
discussion of this until we have set up the tools to work away from
the chiral limit.

The method has been extended, in the chiral limit, to other
bilinears. On-shell improvement requires addition of all dimension
four operators with appropriate quantum numbers~\cite{alpha}\footnote{%
Our convention for the tensor is 
$T_{\mu\nu}^{(jk)}(x)=\bar \psi^j(x) i \sigma_{\mu\nu} \psi^k(x)$
with $\sigma_{\mu\nu}=\frac12 i [\gamma_\mu,\gamma_\nu]$.
}
\begin{eqnarray}
S^{(jk),I} &=& S^{(jk)}
\\
V_\mu^{(jk),I} &=& V_\mu^{(jk)} + a c_V \partial_\nu T_{\mu\nu}^{(jk)}
\label{eq:improvedV}
\\
T_{\mu\nu}^{(jk),I} &=& T_{\mu\nu}^{(jk)} + a c_T \left[
\partial_\mu V_\nu^{(jk)} - \partial_\nu V_\mu^{(jk)} \right]
\,,
\end{eqnarray}
where $j\ne k$.
Each of these operators also has an associated normalization
constant. The improvement constants $c_V$ and $c_T$ can be
determined by enforcing appropriate axial Ward identities in the
chiral limit, making use of the previous determination of the improved
axial current. In particular, enforcing that the axial variation of
$V_\mu$ is proportional to $A_\mu$ determines $c_V$---see
Refs.~\cite{GuagnelliSommer,Martinelli,WIPLB} for
particular methods---while enforcing that $T_{\mu\nu}$ rotates into
other components of itself determines
$c_T$~\cite{Martinelli,WIPLB}.  

Up to this point we have considered only flavor off-diagonal bilinears,
i.e. those with flavor indices satisfying $j\ne k$. 
For the subsequent analysis we will need to consider also
diagonal flavor non-singlet operators (as well as flavor singlets).
A convenient common notation for all non-singlet bilinears is
$\tr(\lambda \CO)$, where the trace is over flavor indices, and
$\lambda$ is one of the eight Gell-Mann matrices for $N_f=3$ (our primary focus),
the Pauli matrices for $N_f=2$, and the appropriate generalization
for $SU(4)$. 
In the chiral limit, where the $SU(N_f)$ flavor symmetry is unbroken,
the improvement of the off-diagonal non-singlets described above carries
over also to the diagonal non-singlets,
with the same improvement coefficients.
Thus, for example, the general improved non-singlet axial current is
\begin{equation}
\tr(\lambda A_\mu)^{I} = \tr(\lambda A_\mu) + a c_A \partial_\mu \tr(\lambda P)
\,,
\end{equation}
for all choices of $\lambda$.

When one moves away from the chiral limit, many new improvement
constants are needed. Consider first the gluon action. Possible higher
dimension gluonic operators are of dimension six, so the only
contribution at $O(a)$ is the original action-density multiplied by
the trace of the quark mass matrix, $M$. This leads to an effective
gluonic coupling constant~\cite{alpha}
\begin{equation}
g_0^2 \longrightarrow \widetilde g_0^2 
= g_0^2 (1 + a\, b_g \tr M /N_f)
\,,
\label{eq:bgdef}
\end{equation}
where $b_g$ is a function of $g_0^2$.
If we work at fixed bare coupling, and vary $\tr M$, then the
effective coupling will vary, as will the lattice spacing, and so
dimensionful quantities will have $O(a)$ contributions proportional to
$a b_g\tr M$. To avoid these, $g_0$ must, in principle, 
be varied as $\tr M$ changes
in such a way that $\widetilde g_0$ is held constant. This requires
determining the improvement constant $b_g$. 
Non-perturbative methods for doing so have
been proposed in Ref.~\cite{Martinelli},
and the
one-loop perturbative result is given in Ref.~\cite{sintsommer}.
%
%

For most of the following discussion we assume that $b_g$ is known,
and that whenever we vary $M$ we do so with
$\widetilde g_0$, and thus $a$, fixed.
In fact, we will find that our method provides an
alternative, independent determination of $b_g$.
Thus one does not need to rely on the methods of
Refs.~\cite{alpha,Martinelli}.

The improvement of quark masses and bilinear operators away from the
chiral limit in an unquenched theory has been discussed previously,
but only in the case of degenerate quarks~\cite{alpha,Martinelli}.
We do not recall
this work here, since we will generalize it in the following to quarks
with non-degenerate masses.

\section{Additional improvement coefficients}
\label{sec:coeffs}

In this section, we describe the new improvement and renormalization
coefficients that are required in order to improve, at $O(a)$, 
flavor singlet
and non-singlet bilinear operators for general values of quark masses.
The reason for the inclusion of flavor singlet bilinears will become
clear later.

Consider first flavor singlet operators in the chiral limit.  Matrix
elements of these operators have ``quark-disconnected'' contractions (in
which the operator connects to external fields through gluons) in
addition to the usual ``quark-connected'' contractions present also for
flavor non-singlet operators. It follows that the improvement
coefficients $c_A$, $c_V$ and $c_T$ will not
 be the same as those for non-singlet operators,
and thus we denote them $\bar c_A$, $\bar c_V$ and $\bar d_T$,
respectively. The differences
begin at two-loop, and thus are of $O(g_0^4)$ barring unforeseen
cancellations.
Furthermore, some of these operators can
``mix'' with purely gluonic operators.  Enumerating the available
gluonic operators, one finds the following forms for on-shell improved
flavor singlet bilinears:\footnote{%
In Ref.~\cite{lat99} we
erroneously concluded that improvement of the tensor bilinear required
the inclusion of an additional gluonic operator, which, however,
vanishes identically.}
\begin{eqnarray}
(\tr A_\mu)^I &=& \tr A_\mu + a\, \bar c_A \partial_\mu \tr P \,;
\label{eq:trAimp}\\
(\tr V_\mu)^I &=& \tr V_\mu + a\, \bar c_V \partial_\nu \tr T_{\mu\nu}
\,;
\label{eq:trVimp}\\ 
(\tr T_\mu)^I &=& \tr T_\mu + a\, \bar c_T \left[
\partial_\mu \tr V_{\nu} - \partial_\nu \tr V_{\mu} \right]
\,;
\label{eq:trTimp}\\
(\tr S)^I &=& a^{-3} e_S + \tr S + 
a\, g_S \Tr(F_{\mu\nu} F_{\mu\nu}) \,;
\label{eq:trSimp}\\ 
(\tr P)^I &=& \tr P + a\, g_P \Tr(F_{\mu\nu} \tilde F_{\mu\nu}) 
\label{eq:trPimp}\,.
\end{eqnarray}
Here we use ``$\tr$'' for the trace over flavor indices, and ``$\Tr$''
for that over color indices.  For later convenience,
it is important that the discretized form
of $\Tr(F_{\mu\nu} F_{\mu\nu})$ is {\em exactly} that combination of Wilson loops
which appears in the gauge action 
[see eq.~(\ref{eq:SGlue}) below],
with an average over loop positions so as to be centered 
on the site where the bilinear is placed.
For $\Tr(F_{\mu\nu} \tilde F_{\mu\nu})$ one can pick any local choice,
e.g. that based on the clover-leaf.

The mixing of $\tr S$ with the identity operator in eq.~(\ref{eq:trSimp})
was overlooked in Ref.~\cite{lat99}.
To improve $\tr S$, 
the coefficient $e_S$ would need to be determined
to an accuracy of $a^4$. 
To calculate hadronic matrix elements, however, one must subtract
disconnected contributions anyway, and this completely removes 
the $e_S$ term. 
Similarly, the  $e_S$ contribution can be canceled when implementing
Ward identities
 by subtracting disconnected contributions, as we discuss
explicitly below.
In this way one can avoid the problem except when calculating
the vacuum expectation value, i.e. the quark condensate. 
To obtain an improved version of the condensate,
one must use another method, as also discussed below.

\bigskip

We now turn to improvement away from the chiral limit. 
We consider explicitly the case of three light dynamical flavors;
the generalizations to two and four flavors are described in
the appendices..
We restrict the discussion to a diagonal bare mass
matrix with positive entries,
$M =\mathrm{diag}(m_1,m_2,m_3)$, thus
avoiding possible phase structure associated with
spontaneous violation of CP~\cite{dashencreutz}.
Treating $M$ as a spurion transforming
in the adjoint representation of the flavor $SU(3)$ group, it is
straightforward to enumerate the allowed improvement terms linear in
quark masses.  For non-singlet operators we find the general form
\begin{equation}
\widehat{\tr(\lambda \CO)} = Z_O\left[
(1 + a\,\bar b_O\,\tr\,M) \tr(\lambda \CO)^I + a\,\frac{1}{2}\,b_O
\tr(\{\lambda,M\} \CO) + a\,f_O\,\tr(\lambda M) \tr \CO \right] \,,
\label{eq:nonsingletO}
\end{equation}
where 
$\CO$ is any of the five bilinears.  
A possible term proportional to
$\tr([\lambda,M]\CO)$ is forbidden by CP invariance---only the
anticommutator of $\lambda$ and $M$ appears.  Note that the
operators appearing in the $O(a)$ corrections on the RHS (right-hand side)
can be chosen
to be improved or unimproved, the difference being of $O(a^2)$.
Here we have left them unimproved for the sake of brevity,
but in some equations below it is more convenient use improved versions.
This choice has no impact on $Z_O$ because the explicit factors of
quark mass do not allow mixing back with the leading operator $\CO$.

There is one subtlety that is overlooked in eq.~(\ref{eq:nonsingletO}).
The flavor-diagonal scalar bilinears (i.e. those for which
the $\lambda$ are diagonal matrices)
can also mix with the identity operator.
For example, the operator
$S^{(jj)}-S^{(kk)}$ mixes with the identity
with a coefficient proportional to $(m_j-m_k)/a^2$.
As for the flavor singlet scalars, however,
this mixing is removed in all but the vacuum
expectation value by subtracting disconnected contributions,
and so we will keep its contribution to
$S^{(jj),I}-S^{(kk),I}$ implicit in the following.

Aside from this subtlety,
there are, for each of the five non-singlet bilinears, three
improvement coefficients $b_O$, $\bar b_O$ and $f_O$, in addition to
the overall normalization in the chiral limit, $Z_O$. 
It is useful to understand the dependence of each of these
quantities on the coupling constant and lattice spacing.
In the chiral limit, the normalization depends both on the bare
coupling and, if the corresponding bilinear has an anomalous dimension,
on $a\mu$ (with $\mu$ the renormalization scale): $Z_O=Z_O(g_0^2,a\mu)$.
As discussed in Ref.~\cite{alpha}, 
away from the chiral limit one must replace $g_0$ with $\widetilde g_0$
of eq.~(\ref{eq:bgdef}) so that, in general, 
$Z_O=Z_O(\widetilde g_0^2,a\mu)$. 
The improvement coefficients do not, however,
 depend explicitly on $a\mu$, and so are functions only of $g_0^2$, 
or equivalently of only $\tilde g_0^2$ at this order of improvement.

To understand the significance of each of the improvement
coefficients, it is useful to consider special cases.  For flavor
off-diagonal operators (which we will also refer to as ``charged''
operators) $f_O$ drops out: 
\begin{equation}
\widehat\CO^{(jk)} = Z_O
\left[1 + a\; \bar b_O \tr\;M + a\; b_O m_{jk}\right] 
\CO^{(jk),I}
\,,
\label{eq:chargedO}
\end{equation}
where $m_{jk} = (m_{j} + m_{k})/2$. The fact that $\bar b_O$
multiplies the trace of $M$ (and thus depends on all three quark
masses) indicates that it arises from mass dependence of quark loops,
and thus begins at two-loop order in perturbation theory,
and is absent in the quenched approximation.  The $b_O$ term, by
contrast, arises from the mass dependence of the valence quark
propagators attached to the operator, and is present also in the
quenched approximation. We have chosen the normalization so as to
match that of the standard form used in quenched applications of
improvement: 
\begin{equation}
\widehat\CO^{(jk)}\big|_{\rm Qu} = Z_O
\left[1 + a\; b_O^Q m_{jk}\right] 
\CO^{(jk),I}
\,.
\label{eq:quenchedchargedO}
\end{equation}
Note that, although $b_O$ and $b_O^Q$ arise from the same underlying
effect, they will differ numerically (for a given choice of $\tilde
g_0^2$) due to mass-independent contributions from quark loops.  Since these enter
first at two loop order, however, one loop results for $b_O$ from
Ref.~\cite{SintWeisz} are valid also for $b_O^Q$.

The $f_O$ term enters into the improvement of flavor diagonal 
(or ``neutral") operators, e.g.
\begin{align}
\widehat\CO^{(jj)} - \widehat\CO^{(kk)} = 
Z_O \big[ (1 &+ a\,\bar b_O\,\tr\,M) (\CO^{(jj),I} - \CO^{(kk),I}) 
 + a\,b_O\,(m_j \CO^{(jj)} - m_k \CO^{(kk)}) \notag \\
&+ a\,f_O\,(m_j-m_k) \tr\,\CO \big] \,.
\label{eq:diagonalO} 
\end{align}
Here, on the left hand side we have made the replacement
\begin{equation}
\widehat{(\CO^{(jj)}\!\!-\!\!\CO^{(kk)})}
= \widehat\CO^{(jj)} - \widehat\CO^{(kk)}
\,,
\label{eq:assocnorm}
\end{equation}
i.e. we have replaced the improved and normalized version
of the operator $(\CO^{(jj)}-\CO^{(kk)})$ with the difference
of the improved and normalized versions of the individual
operators. Similarly, in the first term on the right hand side,
we have used
\begin{equation}
(\CO^{(jj)}-\CO^{(kk)})^I = \CO^{(jj),I} - \CO^{(kk),I}
\,.
\label{eq:associmp}
\end{equation}
One might be concerned that there is a subtlety hidden in these
replacements, since the individual operators $\CO^{(jj)}$ contain
flavor singlet parts and, as discussed further below, flavor singlet
and non-singlet operators can have different anomalous dimensions.
In fact, the same issue arises in the continuum. We resolve it
by simply defining the individual operators as appropriate sums of
the flavor singlet and non-singlet parts, e.g. with two flavors
\begin{equation}
\widehat\CO^{(11)} \equiv \frac12\left[
\widehat{(\CO^{(11)}\!\! -\!\! \CO^{(22)})}
+ \widehat{(\CO^{(11)}\!\! +\!\! \CO^{(22)})} \right]
\,,
\end{equation}
and similarly for the improved operators. Then the relations
in eqs.~(\ref{eq:assocnorm}) and (\ref{eq:associmp}) are identities.

Returning to eq.~(\ref{eq:diagonalO}), it is clear
that the $f_O$ term arises from quark-disconnected contractions
of the operator, because it is only through such 
contractions that mixing with the operator $\tr\,\CO$, which contains all flavors,
can arise. From this we learn that $f_O$ appears first at two-loop order.
We remark that $f_O$ is
also present in improvement of quenched bilinears if one considers
flavor-diagonal non-singlets and keeps disconnected contractions. As
far as we know, no such calculations have been done to date.

The presence of an extra improvement coefficient in the diagonal non-singlet
operators as compared to the off-diagonal non-singlets leads to the
following apparent paradox. 
A non-singlet vector rotation ($\delta_V \psi_j=\psi_k$
and $\delta_V\bar\psi_k = - \bar\psi_j$) transforms off-diagonal operators
into diagonal ones: $\delta_V \CO^{(kj)}=\CO^{(kk)}-\CO^{(jj)}$.
Thus the corresponding Ward identity 
[given explicitly in eq.~(\ref{eq:3ptVWI})]
should be enforced for improved
operators on the lattice up to $O(a^2)$ corrections.
Assuming that $\widehat{\delta_V S}$ is known, this identity
relates an operator with two improvement coefficients ($\bar b_O$ and $b_O$)
to an operator with one more improvement coefficient ($f_O$).
Thus it seems to imply that $f_O$ is not independent.
This would be paradoxical, because $f_O$ is allowed in the first place
by the vector symmetry which is subsequently leading to the constraint.
This paradox is resolved by analyzing the possible contact
terms in Appendix~\ref{app:paradox}.

\bigskip

Previous discussions of improvement in the unquenched theory have
considered only degenerate quarks~\cite{alpha}. 
To make contact with the notation used in these papers, 
we note that the general result (\ref{eq:nonsingletO}) reduces to
\begin{equation}
\widehat{\tr(\lambda \CO)}\big|_{\rm UnQ,\ degen} = Z_O\left[
1 + a\,(b_O + N_f \bar b_O)\,m \right] \tr(\lambda \CO)^I \,,
\label{eq:unquenchednonsingletO}
\end{equation}
for degenerate quarks. 
For charged bilinears, this has the same form as used in Refs.~\cite{alpha},
except that what we call $b_O+N_f\bar b_O$ was denoted simply $b_O$ in
those works. We prefer our notation because of
its connection with the quenched improvement constants and because it
is more easily generalizable to non-degenerate quarks.
We note that at one-loop order the difference in notation is immaterial
because, as noted above, $\bar b_O$ vanishes at this order.

Next we consider the mass-dependent improvement coefficients needed
for flavor singlet operators. We find that the general form consistent
with flavor symmetry is
\begin{equation}
\widehat{\tr\;\CO} = Z_O r_O \left[
(1 + a\,\bar d_O\,\tr\,M) \tr(\CO)^I + a\,d_O\,\tr(M \CO) \right] \,.
\label{eq:singletO}
\end{equation}
There are only two mass-dependent improvement coefficients for each
bilinear (as opposed to the three needed for non-singlets), but there
is an additional normalization factor, $r_O$, appearing in the chiral
limit.  The latter arises because the normalizations of singlet and
non-singlet operators differ, since the former can have quark-disconnected
contractions.  Rather than introduce a new normalization constant
$\bar Z_O$, we have parameterized this effect with the ratio $r_O=\bar
Z_O/Z_O$.  For the axial current ($\CO=A_\mu$) the anomalous dimension
of the singlet and non-singlet operators differ (the former beginning
at two-loop order, while the latter vanishing to all orders), 
so that $r_A$ must
depend explicitly on $\ln a$ in addition to the usual dependence
on $\widetilde g_0^2$.
For the four other bilinears the singlet and
non-singlet anomalous dimensions are the same (the anomalous
dimensions vanish for all vector currents, and the quark-disconnected 
loop diagrams for $S$, $P$ and $T$ vanish by chirality in the continuum).
Thus $r_S$, $r_P$, $r_V$ and $r_T$ do not depend explicitly
on $\ln a$, and are functions only
of the effective coupling constants, just like the improvement coefficients.
See Ref.~\cite{Testa} for a more thorough discussion of this point.

%
%
%

\bigskip
We also need the expressions for $O(a)$ improved quark masses in
terms of the bare quark masses. Using flavor symmetry
and the constraint that the singlet and non-singlet
mass combinations vanish at the same bare mass~\cite{Testa},
we find
\begin{eqnarray}
\widehat{\tr\lambda M} &=& Z_m \left[
(1 + a\,\bar b_m \tr\,M) \tr\lambda M + a\,b_m \tr(\lambda M^2)
\right]
\,, \label{eq:nonsingletM}
\\
\widehat{\tr\;M} &=& Z_m r_m \left[
(1 + a\,\bar d_m \tr\,M) \tr\,M + a\,d_m\tr(M^2) \right]
\,. \label{eq:singletM}
\end{eqnarray}
The notation is analogous to that used for the bilinears.  Note that
there is no separate ``$f_O$-like'' term in (\ref{eq:nonsingletM}),
since such a term, proportional to $\tr(\lambda M)\tr(M)$, can be
absorbed into the $\bar b_m$ term. This reduction in the number of
improvement coefficients will play an important role in the next section, where
we will see how these constants are related to those needed to improve
the scalar bilinear.  The overall constant $Z_m$ is scale-dependent,
but all other constants, including $r_m$~\cite{Testa}, are not.

Since we restrict ourselves to a diagonal mass matrix, the result
(\ref{eq:nonsingletM}) is non-trivial only for diagonal $\lambda$'s.
Taking appropriate linear combinations with
eq.~(\ref{eq:singletM}), one can obtain the result for
individual masses:
\begin{align}
\hat m_j = \hat M_{jj} = Z_m\Bigg\{
\bigg[ m_j &+ (r_m - 1) \frac{\tr M}{N_f} \bigg]
 + a\bigg[ b_m m_j^2 + \bar b_m m_j \tr M \notag \\
&+ (r_m d_m - b_m) \frac{\tr(M^2)}{N_f} 
 + (r_m \bar{d}_m - \bar b_m) \frac{(\tr M)^2}{N_f} \bigg] \Bigg\} \,. 
\label{eq:singleM}
\end{align}
From the $O(1)$ terms we see that $\hat m_j$ vanishes if all bare
masses vanish together. This is by construction. For
non-degenerate positive quark masses, however, 
$\hat m_j$ does not vanish when $m_j=0$. This is not an effect
which vanishes linearly in $a$,
since it remains true even when $O(a)$ terms are dropped, because of the contribution
proportional to $(r_m-1)\tr\,M$. 
In particular, this effect implies that, at fixed gauge coupling, 
the pion becomes massless at a value of the bare up and down quark
hopping parameters (assumed degenerate) which
depends (linearly) on the strange quark mass.
This is similar to the well-known result that the partially
quenched critical hopping parameter differs from the 
fully unquenched value (as discussed, e.g., in Ref.~\cite{Rakowlat04}).

It is useful to consider which of these constants survive in the
quenched approximation.  The constants $\bar b_m$ and $\bar d_m$ do
not, since they are produced by quark loops, as shown by the fact that
they multiply $\tr(M)$.  Similarly, the difference of $r_m$ from
unity, and the difference between $d_m$ and $b_m$, arise from
insertions of the mass on quark loops.  Thus in the quenched
approximation one only needs two constants: $Z_m^Q$ and
$b_m^Q=d_m^Q$. One way of seeing this is to note that $\hat m_j$ can
only depend on $m_j$ in the quenched approximation, which forbids all
but the first and third terms in eq.~(\ref{eq:singleM}).

In summary, moving from degenerate to non-degenerate quarks and
considering singlet as well as non-singlet operators requires the
introduction of a large number of additional improvement coefficients.
All except for $Z_S Z_P$, $Z_T$, $Z_m$ and $r_A$ are scale independent
functions of the effective coupling alone,\footnote{%
$Z_S/Z_P$ is scale independent, while $Z_S Z_P$ is scale dependent.}  
and for these scale-independent quantities
there is no apparent obstacle to their
determination using Ward identities.  Indeed, as we show in the next
section, nearly all can be determined in this way.

\section{Determining coefficients using Ward identities}
\label{sec:WI}

In this section we explain in detail how one can generalize previous
methods to determine most of the new improvement and normalization
coefficients.  We organize our approach into four steps. This is partly
as an explanatory aid, but also because some parts of the later steps
rely on results from the earlier ones. There remains, however,
considerable freedom in the ordering of parts of some steps.

In summary, the four steps are the following.
First, we enforce vector charge conservation,
which determines most of the improvement coefficients, and
the normalization constant, of the vector bilinear. 
It turns out that
this is the only use to which we can put vector Ward identities. 
In Ref.~\cite{lat99} we had claimed otherwise, but this turns
out to be incorrect because of overlooked contact terms.
We explain this point in appendix~\ref{app:VWI}. 

The second step is to relate the improvement and normalization
coefficients for the quark masses to those for scalar bilinears. This
allows us to use, in step three, the simplest axial Ward identities such
as eq.~(\ref{eq:2ptNSaxial}) (``two point Ward identities''), to
determine combinations of coefficients for axial, pseudoscalar and
scalar bilinears.  Finally, in the fourth step we enforce axial Ward
identities in which the axial variation occurs in a region including
other operators (``three point Ward identities'').  Here we have to
deal with contact terms.

\begin{table}[htbp]
\caption{Relations between renormalization and improvement
  coefficients for masses and scalar bilinears. LO and NLO indicate
  leading and next-to-leading order in quark masses.
Results are valid for $N_f=3$ and $4$. They hold also for $N_f=2$
if $b_S$ and $b_m$ are set to zero.
} 
\label{tab:relnI}
\begin{center}
\begin{tabular}{ p{2in} p{4in}  }
Order in $M$ & Relationship or constraint
\\ \hline
LO & $Z_S = 1/Z_m$, $r_S = 1/r_m$, $g_S = b_g/(2 g_0^2)$ \\
NLO & $b_S = - 2 b_m$, $\bar b_S = - \bar b_m$,
$N_f f_S = 2 (b_m - d_m)$, \newline
$d_S = b_S + N_f \bar b_S$,
$d_S+ N_f \bar d_S = -2 (d_m + N_f \bar d_m)$\\
\hline
\end{tabular}
\end{center}
\end{table}


\begin{table}[htbp]
\caption{Normalization and improvement coefficients determined
using various Ward identities (which are denoted schematically)
for $N_f=3$.
LO and NLO indicate leading and next-to-leading order in quark masses.
For completeness, we indicate which 
Ward identities determine $c_{SW}$, $c_V$, $c_A$ and $c_T$.
For the last nine Ward identities (those below the double line), 
we assume that the on-shell improved
flavor non-singlet axial variation of the action has been determined
using the previous identities (i.e. those above the double line). 
The notation ``Not new'' in the final
two lines indicates that these two identities are equivalent,
in the chiral limit, to those considered previously.}
\label{tab:summI}
\begin{center}
\begin{tabular}{ p{1.95in} p{1.2in} p{3.1in} }
Ward identity & LO& NLO
\\ \hline
$\langle H | \sum_{\vec x} \hat V_4^{(jj)} | H\rangle = Q_H^j$ 
&
$Z_V$, $r_V$ 
& 
$b_V$, $\bar b_V$, $f_V$, $d_V$, $\bar d_V$ 
\\ \hline
$\partial_\mu \hat A_\mu^{(jk)} = (\hat m_j + \hat m_k) \hat P^{(jk)}$
\newline and \newline
$\partial_\mu (\hat A_\mu^{(jj)} - \hat A_\mu^{(kk)})$ \newline
$= 2 \hat m_j \hat P^{(jj)} - 2\hat m_k \hat P^{(kk)}$ 
&
$Z_m Z_P/Z_A$,\ $r_m$ \newline
$r_P$,\ $g_P$ \newline
[$c_{SW}$,\ $c_A$]
&
$b_A$,\ $f_A$, \ \
$b_P - 2 b_m$,\ \
$b_m + 2 r_m d_m$ \newline
$(1 + 2 r_m)^2 b_m + 6 r_m (\bar b_m- \bar d_m)$\newline
$2 (2+ r_m) b_m + 3(\bar b_P - \bar b_A + \bar b_m)$ \newline
$b_P + 2 r_P d_P$,\ \
$(2+r_P) b_P + 6 f_P$\newline
$r_P(2+r_P) (\bar b_P - \bar d_P)
- (1 + 2 r_P + 3 r_m r_P) f_P$
\\ \hline
$\delta_A^{(ij)} T^{(jk)} = T^{(ik)} $
& $Z_A$, [$c_T$] 
& $b_T$,\ $3\bar b_A - b_A(r_m-1)$
\\
$\delta_A^{(ij)} V^{(jk)} = A^{(ik)}$\newline
and $V \leftrightarrow A$
& $Z_V$,\ $Z_A^2$, [$c_V$] 
& $b_A + b_V$,\  $3 \bar b_A - b_A(r_m-1)$\newline
$6(\bar b_A - \bar b_V) + (b_A- b_V)(2+r_m)$
\\
$\delta_A^{(ij)} P^{(jk)} = S^{(ik)}$\newline
and $P \leftrightarrow S$
& $Z_S/Z_P$,\ $Z_A^2$ 
& $b_S + b_P$,\ $3 \bar b_A - b_A(r_m-1)$\newline
$6(\bar b_P - \bar b_S) + (b_P- b_S)(2+r_m)$
\\ \hline\hline
$\delta_A^{(ij)} \tr A = \delta_A^{(ij)} \tr V = 0$ &
$\bar c_A$,\ $\bar c_V$ &
$ d_A$,\ $ d_V$ \\
$\delta_A^{(ij)} \tr P = 2 S^{(ij)}$ &
$Z_Pr_P/Z_S$, $g_P$ &
$d_P$,\ $3(\bar b_S - \bar d_P) - b_S (r_m-1)$\\
$\delta_A^{(ij)} \tr T = 2 T^{(ij)}$ &
$r_T$, $\bar c_T$ &
$d_T$,\ $3(\bar b_T - \bar d_T) - b_T (r_m-1)$\\
$\delta_A^{(ij)} \tr S = 2 P^{(ij)}$ &
$Z_Sr_S/Z_P$, $g_S$ &
$d_S$,\ $3(\bar b_P - \bar d_S) - b_P (r_m-1)$\\
$\delta_A^{(ij)} T^{(ji)} = T^{(ii)} + T^{(jj)}$ &
$r_T$,  $\bar c_T$, [$c_T$]&
$b_T$, $d_T$, $r_T(\bar d_T - \bar b_T) - r_m f_T$\\
$\delta_A^{(ij)} S^{(ji)} = P^{(ii)} + P^{(jj)}$  \newline 
and $S \leftrightarrow P$ 
& $Z_P/Z_S$,\ $r_P$,\ $r_S$ \newline $g_P$,\ $g_S$
& $b_P + b_S$, $2 r_P d_P+b_P$, $2 r_S d_S+b_S$ \newline
$6(\bar b_P-\bar b_S) + (2+r_m)(b_P-b_S)$ \newline
$3 r_P(\bar d_P - \bar b_S) + 
r_P (d_P-b_S) - r_m (3 f_P +b_P - r_P b_S)$\newline
$3 r_S(\bar d_S - \bar b_P) + r_S (d_S-b_P) - 
r_m (3 f_S +b_S- r_S b_P)$
\\
$\delta_A^{(ij)} A^{(ji)} = V^{(jj)} - V^{(ii)}$  &
Not new &
$\bar b_V$,\ $b_V$,\ $f_V$ \\
$\delta_A^{(ij)} V^{(ji)} = A^{(jj)} - A^{(ii)}$  &
Not new &
$\bar b_A$,\ $b_A$,\ $f_A$
\\
\hline
\end{tabular}
\end{center}
\end{table}

\newpage

We discuss the $N_f=3$ theory in this section,
summarizing the results in Tables~\ref{tab:relnI} and
~\ref{tab:summI}. The former collects the relationships between
constants, while the latter shows which can be obtained from
which Ward identities. 
The generalizations to $N_f=4$ and $N_f=2$ are discussed,
respectively, in appendices~\ref{app:fourflavors} and \ref{app:twoflavors}.

\subsection{Vector Ward identities}

We use the standard method of enforcing the correct normalization
of the vector charge,
\begin{equation}
\langle H | \widehat\CQ^{j}(\tau) | H\rangle = 
Q_H^j \langle H|H \rangle\,,\qquad\qquad
\widehat\CQ^{j}(\tau) = \sum_{\vec x} \hat V_4^{(jj)}(\vec x, \tau)
\,, \label{eq:vectorWI}
\end{equation}
where $Q_H^j$ is the j'th quark number of hadron H, for a
convenient set of hadrons. 
In fact, one does not have to project onto
a single hadron---any linear combination created by an operator with
a given j'th quark number will work.  As we now show, this method can
determine all improvement and renormalization coefficients for the
vector current, with the exception of $c_V$ and $\bar c_V$.  These
two are excluded because the operators they multiply vanish at
zero spatial momentum.

To see that all other coefficients can be determined we use
eqs.~(\ref{eq:diagonalO}) and (\ref{eq:singletO}) with $\CO=V_4$ at
zero spatial momentum. 
Consider first degenerate quarks, so that
\begin{align}
\widehat\CQ^j - \widehat\CQ^k &= 
Z_V\, [1 + a (3 \bar b_V + b_V)m]\, (\CQ^j - \CQ^k) \,,
\label{eq:diagonalOdegen} \\
\widehat{\tr\CQ} &= Z_V r_V
\,[1 + a (3\bar d_V+d_V)m]\, \tr\CQ \,,
\label{eq:singletOdegen}
\end{align}
where $\CQ^j=\sum_{\vec x} V_4^{(jj)}$ is the bare charge operator.
We have dropped the superscript $I$ since
the $c_V$ and $\bar c_V$ terms do not contribute.
Enforcing the normalization of these two charges for two
or more values of the common quark mass determines $Z_V$, 
$3 \bar b_V + b_V$, $r_V$ and $3 \bar d_V+d_V$.
Note that for the singlet charge operator one must use baryonic states,
so that the total charge is non-zero.
To obtain the remaining mass-dependent improvement coefficients
non-degenerate quarks are required. 
It is sufficient, however,
to work with a ``2+1'' flavor theory, i.e. one in
which the up and down quarks are degenerate 
(with mass $m_1$) but have a different
mass than the strange quark ($m_3$). In this case, 
the improvement term for the flavor non-singlet charge 
$\widehat\CQ^{1}-\widehat\CQ^{3}$ is proportional to
\begin{equation}
a\left\{ 
\left[(2 \bar b_V+b_V) m_1 + \bar b_V\, m_3 \right] \CQ^{1}
-
\left[2 \bar b_V m_1 + (\bar b_V+ b_V) m_3 \right] \CQ^{3}
+ \,f_V\,(m_1-m_3) \tr\CQ \right\}\,,
\end{equation}
as can be seen from (\ref{eq:diagonalO}).
Thus by varying $m_1$ at fixed $m_3$, and considering different
hadrons so that the contributions of the independent operators
$\CQ^{1}$, $\CQ^{3}$ and $\tr\CQ$ vary, one can determine $b_V$,
$\bar b_V$ and $f_V$ separately.\footnote{%
Determination of $b_V$ and $\bar b_V$ alone can be done
using $\CQ^{1}-\CQ^{2}$, for which the $f_V$ term vanishes
in the $2+1$ theory.} 
Similarly, the improvement term for the singlet charge is
proportional to
\begin{equation}
a \left\{ (2\bar d_V +d_V) m_1 (\CQ^{1}+\CQ^{2})
+  (\bar d_V + d_V) m_3 \CQ^{3} \right\}
\,,
\end{equation}
so that $d_V$ and $\bar d_V$ can be disentangled by varying $m_1$.

In practice, $Z_V$ and $3\bar b_V+b_V$ have been computed for two flavors using
this method~\cite{Bakeyev:2003ff} 
(although recall that the latter combination is referred
to as $b_V$ in Ref.~\cite{Bakeyev:2003ff}).
Determination of $b_V$ and $\bar b_V$ separately should be relatively
straightforward, since quark-disconnected contractions are not required.
By contrast, the determination of
$r_V$, $\bar d_V$, $d_V$ and $f_V$ requires such contractions, 
and will thus be more challenging in practice.\footnote{%
One can show that the quark-disconnected diagrams give vanishing
contributions if one uses the charge built out of the conserved vector 
current on the lattice. We can see no argument to extend this result
to choices of the bare current which are not exactly conserved.
}


Vector Ward identities cannot be used to determine any of the other
normalization or improvement coefficients, essentially because the
vector symmetries are not broken by the discretization.  We discuss
this further in Appendix~\ref{app:VWI}, because in some cases it is not
immediately obvious why vector Ward identities cannot be used.
Indeed, in Ref.~\cite{lat99}, we argued that it was possible to use
such identities, and it is instructive to see the flaw in our
argument.

\subsection{Relating improvement of mass and scalar bilinear}

As is well known, the anomalous dimensions of the quark mass and
scalar bilinear are equal in magnitude but opposite in sign, 
and it is conventional and convenient to
choose their renormalization constants to be the inverse of one
another, $Z_S Z_m=1$. While not strictly necessary (the right hand
side could be a constant other than unity), this choice implies that
derivatives with respect to quark mass give rise to insertions of the
(space-time integral) of the scalar bilinear both for bare and
renormalized quantities.  In particular, it implies that the useful
result
\begin{equation} 
\frac{\partial m_{H}}{\partial {m}_{j}}\bigg\arrowvert_{m_{k\neq j}}
=
\frac{V\; \langle H | {S}^{(jj)} | H \rangle}{\langle H | H\rangle}
\label{eq:sigmatermcont}
\end{equation} 
(where $|H\rangle$ is the state containing an arbitrary
hadron of mass $m_H$ at rest, $V$ is the spatial volume,
and $k$ runs over the flavors different from $j$)
holds for both bare and renormalized quantities.

In this section we apply the condition (\ref{eq:sigmatermcont}) as a
constraint not only in the continuum limit but also for non-vanishing
lattice spacing. The argument for doing so is straightforward:
the quantities appearing in the 
relation---$\langle H | {S}^{(jj)} | H \rangle$,
$m_H$ and $m_j$---are physical quantities that should
be improved with appropriate choices of the improvement
coefficients introduced above.  Thus if the relation
(\ref{eq:sigmatermcont}) holds in the continuum limit it should
receive no $O(a)$ corrections once improvement has been implemented.
Perhaps surprisingly, this simple relation leads to a number of
non-trivial constraints.

The precise relation we enforce is\footnote{%
We stress that all the lattice quantities appearing in this relation
are defined to have the same dimension as their continuum counterparts,
e.g. $a m_H$ is the hadron mass in lattice units.}
\begin{equation} 
\frac{\partial m_{H}}{\partial \widehat{m}_{j}}
\bigg\arrowvert_{\widehat{m}_{k \neq j},a} =
\frac{V \; \langle H | \widehat{S}^{(jj)} | H \rangle}{\langle H | H \rangle}
\,. 
\label{eq:SMR}
\end{equation} 
Here we have been specific about the meaning of the partial derivative
in the presence of the regulator.  It should be taken with the
regulator---here the lattice spacing $a$---fixed, so that the
relation survives in the limit that the regulator is removed.  In order to
match the formal result (\ref{eq:sigmatermcont}) the derivative should
also be taken with the other {\em improved} (rather than bare)
masses held fixed. Finally, note that the matrix element of
$\hat S$ is, as usual, the connected matrix element, so that the
part of $\hat S$ proportional to the identity 
[the $e_S$ term in eq.~(\ref{eq:trSimp})] does not contribute.

Despite the fact that it involves the {\em a priori} unknown quantities
$\langle H | \widehat{S}^{(jj)} | H\rangle$,
this relation is useful because we do
know how hadron masses depend on bare parameters. 
In particular, because the fermion action 
depends on the bare quark masses only as
\begin{equation}
S_{lat, F} =
\sum_x \sum_j (a m_j) \left(a^3 S^{(jj)}[x]\right) + \dots
\end{equation}
(even after improvement by the addition of the clover term) 
one can show that~\cite{scalardensities} 
\begin{equation} 
\frac{\partial (a m_{H})}{\partial (a m_{j})} 
\bigg\arrowvert_{(a m)_{k \neq j},g_0^2} = 
\frac{ L^3 \langle H | (a^3 S^{(jj)}) | H \rangle}{\langle H | H\rangle} 
=
\frac{ V \langle H | S^{(jj)} | H \rangle}{\langle H | H\rangle} 
\,.
\label{eq:dmHdmj}
\end{equation}
In the derivative on the LHS (left-hand side) the {\em bare} coupling and
the other bare quark masses {\em in lattice units} are held fixed.
As mentioned above, it is the connected matrix element
of the scalar density which appears on the RHS.
Similarly, one finds
\begin{equation} 
-2 g^{4}_{0} \frac{\partial (a m_{H})}
{\partial g^{2}_{0}} \bigg\arrowvert_{a m_{j}} =
\frac{a V \langle H | \Tr(F_{\mu \nu}F_{\mu \nu})| H \rangle}
{\langle H|H\rangle} 
\,.\label{eq:dmHdg}
\end{equation}
where we {\em define} the discretized field strength from the 
specific form of the gluon action being used:
\begin{equation}
S_{lat, G} \equiv a^4 \sum_x \frac{1}{2 g_0^2}
\Tr(F_{\mu \nu}F_{\mu \nu})(x) \,. 
\label{eq:SGlue}
\end{equation}
In this way, our expressions hold for any choice of gluon action.

To relate the desired derivative in eq.~(\ref{eq:SMR}) to those
which we know, (\ref{eq:dmHdmj}) and (\ref{eq:dmHdg}),
we proceed in two stages.
First, we relate derivatives with respect to
improved masses to those with respect to bare masses,
using the properties of Jacobians:
\begin{align} 
\label{eq:SMRrhs}
\frac{\partial m_{H}}{\partial \widehat{m}_{j}} 
\bigg\arrowvert_{\widehat{m}_{k},\widehat{m}_{l},a} 
&= \frac{\partial (m_{H}, \widehat{m}_{k}, \widehat{m}_{l})}
{\partial (\widehat{m}_{j}, \widehat{m}_{k}, \widehat{m}_{l})} 
\notag \\
&= \frac{\partial (m_{H}, \widehat{m}_{k}, \widehat{m}_{l}) 
         / \partial (m_{j},m_{k},m_{l})}
{\partial (\widehat{m}_{j}, \widehat{m}_{k}, \widehat{m}_{l}) 
         / \partial (m_{j},m_{k},m_{l})} \,, 
\end{align}
where $k$ and $l$ are the two flavor indices not equal to $j$.
Here all derivatives are at fixed $a$, or equivalently at fixed 
$\tilde g_0^2$. This means that the derivatives such as
$\partial \widehat{m}_k/\partial m_j$ can be straightforwardly
evaluated using eq.~(\ref{eq:singleM}).
The renormalization constants $Z_m$ and $r_m$
are functions of $\tilde g_0^2$ and $a$, and thus are fixed.
The same is true, to the order in $a$ that we are working,
of the improvement constants $b_m$ etc., since they can
equally well be considered functions of $g_0^2$ or $\tilde g_0^2$.

Derivatives such as $\partial m_H/\partial m_j$ cannot
yet be evaluated since they are taken at fixed $a$.
Using eq.~(\ref{eq:bgdef}), one can relate them to
the derivatives we know from eqs.~(\ref{eq:dmHdmj}) and (\ref{eq:dmHdg}):
\begin{equation}
\frac{\partial m_H}{\partial m_j}\bigg\arrowvert_{m_{k\ne j},a} 
=
\frac{\partial (a m_H)}{\partial (a m_j)}\bigg\arrowvert_{(am)_{k\ne j},a}
=
\frac{\partial (a m_H)}{\partial (a m_j)}\bigg\arrowvert_{(am)_{k\ne j},g_0^2} 
- \frac{g_0^2 b_g}{N_f}
\frac{\partial (a m_H)}{\partial {g_0^2}}\bigg\arrowvert_{am_j} 
\,.
\label{eq:dmHdmII}
\end{equation}
Putting things together we obtain
an expression for the right-hand side of eq.~(\ref{eq:SMR})
in terms of matrix elements of
the bare scalar density and the gluon field strength,
and the improvement and normalization constants for quark masses.
The left-hand side of eq.~(\ref{eq:SMR}) can be expanded,
using eqs.~(\ref{eq:diagonalO}) and (\ref{eq:singletO}),
in terms of the same matrix elements,
but with the coefficients being the improvement and normalization
constants for the scalar bilinear.
Matching the coefficients on the two sides gives, after tedious algebra,
the relations quoted in Table \ref{tab:relnI}.

We do not go through the details of this calculation in the general case,
but do display the subset of the argument that uses
only degenerate quarks. In this case, eqs.~(\ref{eq:SMRrhs})
and (\ref{eq:dmHdmII}) simplify to
\begin{align}
\frac{\partial m_H}{\partial \widehat m}\bigg\arrowvert_{a} 
&= 
\frac{\partial(a m_H)/\partial(a m)|_{a}}
     {\partial{\hat m}/\partial{m}|_{a}} \\
&=\frac{\partial(a m_H)/\partial(a m)|_{g_0^2}
- g_0^2 b_g \partial(a m_H)/\partial(g_0^2)|_{am}}
     {\partial{\hat m}/\partial{m}|_{a}} \\
&= 
\frac{V \langle H |\tr S + a (b_g/2g_0^2) \Tr(F_{\mu\nu}F_{\mu\nu})| H\rangle}
     {Z_m r_m [1 + 2 a (3 \bar d_m + d_m) m]}
\,,
\end{align}
where in the last step we have assumed $\langle H|H\rangle=1$.
This should be equated with the improved matrix element 
\begin{equation}
V \; \langle H | \tr \widehat{S} | H \rangle
= V
\langle H |
Z_S r_S \left\{[1 + a (3 \bar d_S + d_S)m ]\tr S 
+ a g_S \Tr(F_{\mu\nu}F_{\mu\nu})\right\}
| H\rangle
\,,
\end{equation}
where, as noted above, the part of $\widehat{\tr S}$ proportional
to the identity operator does not contribute.
We conclude that $Z_S r_S=1/(Z_m r_m)$,
$3 \bar d_S + d_S = - 2(3 \bar d_m + d_m)$
and $b_g = 2 g_0^2 g_S$, which are a subset of the
results in Table~\ref{tab:relnI}.

We find that the quenched relations $Z_m Z_S = 1$ and $b_S=- 2 b_m$
(Ref.~\cite{SintWeisz}) continue to hold (although the constants
themselves will differ from their quenched values),
and that there are generalizations for some of the new constants 
that appear ($\bar b_S=-\bar b_m$, etc.).
There are two particularly interesting, and perhaps unexpected, results.
First, there is a constraint on the scalar improvement
coefficients ($d_S = b_S + 3 \bar b_S$). This arises because,
as noted above, there is one less improvement constant needed for
quark masses than for the scalar bilinear (i.e. there is no
independent $f_m$ term).
Second, the relation $b_g = 2 g_0^2 \;g_S$ provides another way
of determining $b_g$, if we determine
$g_S$ using Ward identities as discussed below.

Given the relations of Table~\ref{tab:relnI}, it is interesting
to determine if there are any products of masses times scalar densities
which maintain their form under improvement. This does {\em not} hold
for the contribution of mass terms to the action itself,
i.e.
\begin{equation}
\sum_j \widehat{m}_j \widehat{S}^{(jj)}\ne \sum_j m_j S^{(jj)}
\,.
\end{equation}
The corrections to this equation in the general case of
non-degenerate quarks are lengthy and uninformative,
so we quote the result only for the case of degenerate quarks
\begin{equation}
\widehat{m}\ \tr\; \widehat{S}\bigg|_{m_i=m_j=m_k}
= m\, \tr\; S +
 a (3\bar d_m + d_m) m^2\, \tr S
+ a g_S m \Tr(F_{\mu\nu} F_{\mu\nu})
\,.
\end{equation}
Here we have dropped the term containing the identity operator
since it does not contribute to matrix elements.
This degenerate quark result illustrates the general point that
there is, at $O(a)$, mixing with other terms in the action,
and thus no reason to expect that each term in the action should be
separately form invariant under improvement.
The one example of form invariance is for the variation of
the action under vector transformations:
\begin{equation}
(\widehat m_j - \widehat m_k) \widehat S^{(jk)} =
(m_j - m_k) S^{(jk)}
+ O(a^2) \,.
\label{eq:mS}
\end{equation}
This follows from the definitions (\ref{eq:nonsingletO})
and (\ref{eq:nonsingletM}) and the relations $Z_S Z_m = r_S r_m =1$,
$b_S=-2b_m$ and $\bar b_S = - \bar b_m$. 
It holds for $N_f=2-4$ (and we suspect for all higher $N_f$ as well).
This relation plays an important role in the discussion
of vector Ward identities in Appendix~\ref{app:VWI}.

\subsection{Two-point axial Ward identities}

In this section we investigate which of the improvement
and normalization coefficients can be determined using
two-point Ward identities such as eq.~(\ref{eq:2ptNSaxial}).
To obtain as much information as possible we need to vary
the quark masses independently (as done in the quenched
theory in Ref.~\cite{Petronzio}) and consider the PCAC relation
for both charged and neutral currents.

We assume that $c_A$ has been determined, 
so that we know $A^{(jk),I}$. 
We can then calculate the Ward identity mass,
\begin{equation}
\widetilde{m}_{jk} 
\equiv \frac{\langle \partial_{\mu} A^{(jk), \,I}_{\mu}(x)\rangle_J} 
{2 \langle P^{(jk)}(x) \rangle_J} \qquad (j\ne k)
\,.
\label{eq:WImassdef}
\end{equation}
Because we have improved the action and the axial current,
$\widetilde{m}_{jk}$ should be independent of $x$ and of the source $J$ up
to corrections of $O(a^2)$, and thus we do not give it any arguments,
nor specify the source.
We imagine choosing a source with a good signal,
varying the quark masses (keeping $\tilde g_0^2$ fixed,
as always) and studying the 
bare quark mass dependence of $\widetilde{m}_{jk}$.
Using the Ward identity (\ref{eq:2ptNSaxial}), we have
\begin{align}
\widetilde{m}_{jk}
&= \frac{1}{2}
(\widehat{m}_{j} + \widehat{m}_{k}) 
\frac{Z_{P}( 1 + a\; \overline{b}_{P}  \tr M  + a\; b_{P}  m_{jk})} 
{Z_{A}( 1 + a\; \overline{b}_{A} \tr M  + a\; b_{A} m_{jk})} 
\,, \\ 
&= \frac{Z_{P} Z_{m}}{Z_{A}} 
\left[ m_{jk} + \frac{\tr M }{3}(r_{m} - 1) 
+ a\left(
 \mathcal{A} m_{jk}^2 
+\mathcal{B} m_{jk} \tr M
+\mathcal{C} \tr(M)^2
+\mathcal{D} \tr(M^2)\right) \right]
\,, \label{eq:mAWI} \\
\mathcal{A} &= b_P - b_A - 2 b_m
\,, \\
\mathcal{B} &= \bar b_P - \bar b_A + \bar b_m + 2 b_m 
                + \frac{r_m-1}{3}(b_P - b_A)
\,, \\
\mathcal{C} &= \frac{r_m-1}{3} (\bar b_P - \bar b_A)
                + \frac{r_m \bar d_m - \bar b_m}{3} - \frac{b_m}{2}
\,, \\
\mathcal{D} &= \frac{2 r_m d_m + b_m}{6}
\,.
\end{align}
From the terms linear in bare quark masses one can extract
$Z_m Z_P/Z_A$ (as in the quenched case) and $r_m$ (absent in
the quenched case). For example, the derivative of
$\widetilde{m}_{jk}$ with respect to $m$ at $M=0$ for degenerate
quarks is $r_m Z_m Z_P/Z_A$, while the derivative with respect to
$m_l$, $l \neq j,\,k$ alone is $(r_m-1) Z_m Z_P/(3 Z_A)$ (a
quantity which vanishes in the quenched theory).
Using the relations in Table~\ref{tab:relnI} one has thus 
determined $Z_P/(Z_A Z_S)$ and $r_S=1/r_m$.

From the quadratic terms one can determine the coefficient
of each of the four linearly independent functions of masses that appear.
Thus we obtain four linear combinations of the eight
constants $b_A$, $\bar b_A$, $b_P$, $\bar b_P$, $b_m$, $\bar b_m$,
$d_m$ and $\bar d_m$ (given that we know $r_m$ from above).
By comparison, in the quenched approximation, where there are
only three constants, $b_A^Q$, $b_P^Q$ and $b_m^Q=d_m^Q$, one
can determine the two combinations 
$b_A^Q-b_P^Q$ and $b_m^Q$~\cite{Petronzio}.

To obtain further information, we generalize
the method by considering flavor-diagonal two-point Ward identities, 
e.g.
\begin{equation} 
\left\langle \partial_{\mu} (\widehat{A}^{(11)}_{\mu}(x) 
- \widehat{A}^{(22)}_{\mu})(x)\right\rangle_J = 
\left\langle 2 \widehat{m}_{1} \widehat{P}^{(11)}(x) - 2 \widehat{m}_{2}
\widehat{P}^{(22)}(x) \right\rangle_J +O(a^2) \, .
\label{eq:2ptDAWI} 
\end{equation}
This introduces several of the new constants present
in the unquenched theory: $f_A$ on the left-hand side, and
$f_P$, $d_P$ and $\bar d_P$ on the right-hand side
(which contains the flavor-singlet $\hat P$).
This corresponds to the fact that, for $m_1\ne m_2$,
there are contractions in which the source $J$ and the axial current
(or pseudoscalar density) are not connected by quark propagators.
In fact, it is only for $m_1\ne m_2$ that this identity gives new
information: for degenerate quarks, and with appropriate sources,
the contractions are exactly the same as those for the
flavor off-diagonal identity (\ref{eq:2ptNSaxial}).

To enforce the Ward identity, we have to adjust the constants so
that the LHS and RHS are equal
for all choices of $x$ and sources $J$, up to $O(a^2)$. 
We cannot simply take their ratio,
as we did for the off-diagonal Ward identity,
since more than one operator
contributes on both sides of the equation, with relative
strengths that we do not know {\em a priori}.
To proceed, it is useful to expand out
the operator appearing in the left-hand side of (\ref{eq:2ptDAWI}):
\begin{align}
\mathrm{LHS} &= Z_{A} 
\left[1 + \frac{(b_{A} + 3 \overline{b}_{A})}{3} \tr M
        + \frac{b_A}{6} \tr(\lambda'_8 M) \right]
\left\{\partial_{\mu} \tr(\lambda_{3} A_\mu)^I
+ a  \mathcal{E}\, \tr(\lambda_{3} M) \right\}
\,, \label{eq:diagAWILHS}  \\
6 \mathcal{E} &=  2(b_{A} + 3 f_{A}) \partial_{\mu} \tr A_\mu 
                +   b_{A} \partial_{\mu} \tr(\lambda'_{8} A_\mu)
\,.
\end{align}
For convenience we have defined $\lambda'_8=\sqrt 3 \lambda_8$.
We can now divide the Ward identity (\ref{eq:2ptDAWI}) 
by the overall function of the masses multiplying
$\partial_\mu \tr(\lambda_3 A_\mu)^I$, so that
the operator on the left-hand side becomes
\begin{equation}
\mathrm{LHS}' \equiv \frac{\mathrm{LHS}}
{Z_A \left[1 + \frac{(b_{A} + 3 \overline{b}_{A})}{3} \tr M
        + \frac{b_A}{6} \tr(\lambda'_8 M) \right]} 
= \partial_{\mu} \tr(\lambda_{3} A_\mu)^I
+ a \mathcal{E}\, \tr(\lambda_{3} M) \,. 
\label{eq:diagAWILHS'}
\end{equation}
Since we have previously determined the improvement
coefficients in the leading order operator in this equation, 
$\partial_\mu \tr(\lambda_3 A_\mu)^I$,
the coefficients
of all other, independent, operators appearing on both sides of the 
rescaled Ward
identity can be determined.
In particular, we immediately see that $b_A+3 f_A$ and $b_A$
can be determined since they multiply independent operators
in $\mathcal{E}$.

We now divide the right-hand side by the same factor,
and split the operator which results into terms linear
and quadratic in quark masses:
\begin{equation}
\mathrm{RHS}' \equiv
\frac{\mathrm{RHS}}
{Z_A \left[1 + \frac{(b_{A} + 3 \overline{b}_{A})}{3} \tr M
        + \frac{b_A}{6} \tr(\lambda'_8 M) \right]} 
=
\mathrm{RHS}'_{I} + \mathrm{RHS}'_{II}
\,.
\end{equation}
We find
\begin{align}
\mathrm{RHS}'_{I} = 
\frac{Z_{m} Z_{P}}{3 Z_A} \bigg\{ &\tr(\lambda_{3} P) 
\left[2 r_m \tr M +  \tr(\lambda'_{8} M)\right] \notag \\
+ &\tr(\lambda_{3} M) \left[ 2 r_{P} \tr P + \tr(\lambda'_{8} P)
+ 2 a r_{P} g_{P} \Tr(F_{\mu\nu} \widetilde F_{\mu\nu}) \right]
\bigg\} \,.
\end{align}
We can determine the coefficients of each independent function
of masses (of which there are three at this order)
multiplying each independent operator.
Thus we can determine $Z_m Z_P/Z_A$ and $r_m$ again,
as well as $r_P$ and $g_P$ for the first time.

The quadratic terms are more complicated. 
There are six independent functions of the masses,
each potentially multiplying three independent operators, 
although the $1\leftrightarrow 2$ antisymmetry allows only
eight independent products.
We find
\begin{align}
\mathrm{RHS}'_{II} = a \frac {Z_{m} Z_{P}}{9 Z_A}
\bigg\{ &\left[ \mathcal{F} \tr(M)^{2} 
              + \mathcal{G} \tr(\lambda_{3}M)^2  
              + \mathcal{H} \tr(M)\tr(\lambda'_{8} M)
              + \mathcal{I} \tr(M^2) \right] \tr(\lambda_3 P) \notag \\ 
+ &\,\left[\mathcal{J} \tr(\lambda_{3} M)\tr(\lambda'_8 M) 
       + \mathcal{K} \tr(\lambda_{3} M)\tr M \right] \tr P \notag \\
+ &\,\left[\mathcal{L} \tr(\lambda_{3} M)\tr(\lambda'_8 M) 
       + \mathcal{M} \tr(\lambda_{3} M)\tr M \right] \tr(\lambda'_8 P)
\bigg\} \,, \\
\mathcal{F} &= 
(2 r_m - 1) (b_P - b_A)
+ 6 r_m(\bar b_{P}- \bar b_A + \bar d_m) + b_m  
=\CA+6\CB+18\CC \,, \\
\mathcal{G} &= \frac{3}{2} (b_A + 2 b_m + 2 r_P d_P) 
\,, \\
\mathcal{H} &= 3( \bar b_P + \bar b_m - \bar b_A)
+ (r_m+1)(b_P-b_A) + 2 b_m 
= 2\CA + 3\CB
\,, \\
\mathcal{I} &= 3(b_P-b_A + 2 r_m d_m - b_m)
=3\CA + 18\CD \,, \\
\mathcal{J} &= 2 b_P + 6 f_P + 2 r_P b_m - r_P b_A
\,, \\
\mathcal{K} &= 6 r_P(\bar b_m + \bar d_P - \bar b_A)
+ 2 r_m(b_P + 3 f_P) + 2 r_P(2 b_m + d_P - b_A)
\,, \\
\mathcal{L} &=b_m + r_P d_P - b_A/2
=\CG/3-b_A/2\,, \\
\mathcal{M} &= 3(\bar b_P + \bar b_m - \bar b_A) + (b_P-b_A+2 b_m) 
+ r_m b_P = \CH + r_m b_A
\end{align}
Thus one can determine, in principle, the eight combinations
$\mathcal{F}-\mathcal{M}$. As indicated, however, only three of
these are independent of the combinations
$\mathcal{A}-\mathcal{D}$ that one can obtain
using the flavor off-diagonal two-point Ward identities.
Thus one can only determine seven combinations of the ten
improvement constants which enter
(we exclude $b_A$ and $f_A$ since we have determined these from the
left-hand side of the present identity).
A particular choice of these seven combinations, is listed
in Table~\ref{tab:summI}.
To further disentangle the coefficients requires  three-point Ward identities,
which we consider below.

As noted above,
implementation of the flavor-diagonal Ward identity necessarily involves
quark-disconnected contractions and thus will be numerically challenging.
Thus it is interesting to know how many combinations of quark masses are needed.
Are simulations in a $2+1$ flavor theory sufficient, or do all three quarks
need to be degenerate? It turns out that a combination of simulations
with degenerate masses (taking three or more values) and $2+1$ simulations
(with at least two values of the light quark mass differing from each of
two choices for the strange quark mass) is sufficient. This allows one
to disentangle all the different linear and quadratic mass dependences
that appear.

\subsection{Three-point axial Ward identities}

We now turn to the enforcement of the chiral transformation properties
of bilinear operators. The methodology is standard~\cite{Bochicchio,alpha,
GuagnelliSommer,Martinelli,WIPLB}; what we add here
is the generalization to non-degenerate masses in the unquenched
theory, and the use of identities involving flavor-singlet components.
Contact terms limit what can be extracted with this method,
and will need to be understood in detail.

We begin with the simplest example, which is that considered
in previous calculations. A non-singlet axial transformation with flavor $(ij)$
is performed in a region of space-time $\mathcal{V}$
which includes a bilinear $\CO^{(jk)}$ with $k\ne i$.
If $\mathcal{O}^{(jk)} = \overline{\psi}^{j} \Gamma \psi^{k}$,
 then it is transformed into 
$\delta \mathcal{O}^{(ik)} =
\overline{\psi}^{i} \gamma_{5} \Gamma \psi^{k}$.
The identity we enforce on the lattice is thus
\begin{equation} \label{eq:AT}
\langle (\delta^{(ij)}_{A}\mathcal{S}) 
\widehat\CO^{(jk)}(y) J^{(ki)}(z) 
\rangle = 
\langle \widehat{\delta \mathcal{O}}^{(ik)}(y) 
J^{(ki)}(z) \rangle + O(a^2) \, ,
\end{equation}
where $\delta_{A}\mathcal{S}$ is the improved and normalized
lattice form of the 
formal variation of the continuum action
\begin{equation} \label{eq:AvarS}
(\delta^{(ij)}_{A} \mathcal{S})= a^4 \sum_{\mathcal{V}} 
 \Big[ (\widehat{m}_{i} + \widehat{m}_{j})
\widehat{P}^{(ij)} - \partial_{\mu} \widehat{A}^{(ij)}_{\mu} \Big] \, ,
\end{equation}
with $\mathcal{V}$ a 4-dimensional subset of the
lattice containing $y$ but not the source at $z$.
(We do not place a ``hat" on $\delta_A\mathcal{S}$ so as to
avoid overloading the notation.)
Note that only quark-connected contractions contribute
to this Ward identity.

Actually, as is well known, the identity~(\ref{eq:AT}) 
cannot be satisfied simply by on-shell improvement,
since the pseudoscalar density appearing in $(\delta_A^{(ij)}\mathcal{S})$
comes into contact with $\CO^{(jk)}$.
Additional off-shell improvement terms are needed,
having the same quantum numbers as the product
$P^{(ij)} \CO^{(jk)}$, and having the appropriate dimension.
Since there is an explicit factor of $\hat m_i + \hat m_j$,
the only such term with the right dimension and symmetries is
${\delta\CO^{(ik)}}$. In previous work, we have used a 
mnemonic for obtaining this contact term, namely that
we can off-shell improve the bilinears by introducing
an additional ``equations-of-motion" operator~\cite{WIPLB}:
\begin{equation}
a  \overline{\psi}^{j} \Big( - \overleftarrow{D \!\!\!\! 
/} + m_{j} \Big) \Gamma \psi^{k} + a 
\overline{\psi}^{j} \Gamma \Big( \overrightarrow{D
\!\!\!\! /} + m_{k} \Big) \psi^{k} 
\,.
\label{eq:EoMops}
\end{equation}
While adequate for discussing the Ward identities (\ref{eq:AT}),
this is potentially misleading for two reasons.
First, the form implies that contact terms between two
operators can be factorized into the contribution of
one operator times that of the other. In fact, there is
no such factorization of contact terms.
Second, for the operators we consider below, there
are several possible contact terms and the mnemonic cannot
be easily generalized.
Thus we do not use this mnemonic further in this paper.
In fact, all we need to know is that there exist possible
contact terms involving the operator $\delta \CO^{(ik)}$
and multiplied by $\widehat m_i+\widehat m_j \propto \widetilde m_{ij} + O(a)$.

A convenient way of using eq.~(\ref{eq:AT}) is to take the ratio
of the two sides having pulled out unknown mass factors.
We first define the improved (but not normalized) variation of the
action by 
\begin{equation}
(\delta_A^{(ij),I}\mathcal{S}) = \frac{(\delta_A^{(ij)}\mathcal{S})}
{Z_A (1 +a \bar b_A \tr M + a b_A m_{ij})}
= a^4 \sum_{\mathcal{V}} 
\left[ 2 \widetilde m_{ij} P^{(ij),I} - \partial_\mu A_\mu^{(ij),I}\right] 
\,.
\end{equation}
This can be determined by calculating
$\widetilde m_{ij}$ using eq.~(\ref{eq:mAWI}), since by assumption
we know $A_\mu^I$.
The ratio we consider is then
\begin{align}
R_O &= \frac{
\langle (\delta^{(ij),I}_{A}\mathcal{S}) 
\mathcal{O}^{(jk),I}(y) J^{(ki)}(z) \rangle }
{
\langle \delta \mathcal{O}^{(ik),I}(y) J^{(ki)}(z) \rangle} \,,
\\
&= \frac{Z_{\delta O}}{Z_A Z_O}
\left[1 + a (\bar b_{\delta O} - \bar b_O - \bar b_A) \tr M
+ a (b_{\delta O} m_{ik} - b_O m_{jk} - b_A m_{ij})\right]
 + a c'_O \widetilde m_{ij} + O(a^2)
\,. 
\end{align}
Here the contact term is included with an
unknown coefficient $c'_O$.\footnote{%
Our convention for $c'_O$ differs from that
defined in Ref.~\cite{WIPLB}.} 
Requiring $R_O$ to be independent of $y$ and $J$ 
in the chiral limit determines $c_V$ and $c_T$,
and we assume this has been done, so that we know all
the $\CO^{(ij),I}$. One also obtains information
on the $Z_O$, as first noted in Ref.~\cite{Bochicchio}.
Away from the chiral limit $R_O$ should be automatically
independent of $y$ and $J$ up to $O(a^2)$,
since there are no additional operators with coefficients
to tune. Note that the contact term plays no role in this regard.
It has the same operator $\delta \CO$ present in both numerator
and denominator, and so is independent of $y$ and $J$ by itself.

Thus, for convenient choices of $y$ and $J$,
evaluating $R_O$ away from the chiral limit allows
one to determine one
combination of improvement coefficients for each of the three
independent linear functions of the quark masses,
except that one of these is not useful as it is 
``contaminated'' by the contact term.
One complication compared to the quenched case is that the
mass dependence of the contact term,
$\widetilde m_{ij}$, is not simply proportional to $m_{ij}$,
but has an additional part proportional to $(r_m-1)\tr M$,
as can be seen from eq.~(\ref{eq:mAWI}).
Because of this, a convenient choice of basis is
$\widetilde m_{ij}$, $m_i-m_j$ and $\tr M$,
using which we find
\begin{align}
R_O 
&= \frac{Z_{\delta O}}{Z_A Z_O}
\bigg\{ 1 + a \left[\bar b_{\delta O} - \bar b_O - \bar b_A + 
(b_{\delta O} - b_O) (2+r_m)/6
+b_A (r_m-1)/3 \right] \tr M
\notag \\
&\qquad
+ a (b_{\delta O} + b_{O}) (m_i - m_j)/4 \bigg\}
+ a \left[ c'_O - (b_{\delta O} - b_{O} + 2 b_A)
  \frac{Z_{\delta O}}{2 Z_{O} Z_P Z_m} \right] \widetilde m_{ij} 
+ O(a^2) 
\,.
\end{align}
The combinations of the
improvement coefficients multiplying $\tr M$ and
$m_i-m_j$ in this equation can, in principle,
be determined. To do so it is sufficient
to use a $2+1$ flavor theory: the coefficient of $\tr M$ can be
determined using $m_i=m_j\ne m_k$, and that of $m_i-m_j$
can then be determined using $m_i\ne m_j=m_k$.
To simply the discussion below, we note that a quick way
of determining the accessible combinations of improvement
coefficients is to set $\widetilde m_{ij}=0$, so that
one can make the substitution $m_{jk}=-\tr M (r_m-1)/3$.
The resulting coefficients of $m_i-m_j$ and $\tr M$ are
those that can be determined.\footnote{%
Note that this is a theoretical device and not a practical
tool. Setting $\widetilde m_{ij}=0$
and considering $m_1-m_2\ne0$ implies
that some quark masses are negative. This is undesirable
in practice due to the possible phase structure at $m\sim 0$.
In practice one would likely need do a fit using positive
quark masses in order to separate the different mass dependencies.
}

Applying this method to the bilinears in turn,
we obtain the results given in Table~\ref{tab:summI}.
These allow the determination of four new quantities:
 $Z_A$, $Z_S/Z_P$, $b_T$ and $\bar b_A$
(where for the latter we have used knowledge of $b_A$ 
and $r_m$ from the two-point Ward identities).
The combinations of $S$ and $P$ improvement coefficients
that are obtained, however, are all related to those obtained from
the two-point Ward identities 
using the relations listed in Table~\ref{tab:relnI}.
Thus we obtain a check of these relations,
but no new information on the constants themselves.

\bigskip
To determine further improvement coefficients we 
consider axial Ward identities involving the transformation
of flavor diagonal bilinears. These have not been considered
previously, and, in particular, they are not needed in the
quenched approximation. They involve quark-disconnected contractions
in an essential way.
The analysis is simplified by the observation
that the Ward identities considered above allow the complete determination
of $\widehat A_\mu^{(ij)}$
and thus of the improved variation of
the action, $(\delta_A^{(jk)}\CS)$, {\em including} its normalization.
Thus the only unknown coefficients appearing in the 
three-point axial Ward identities we consider below are those in
the operator $\CO$ and its axial variation $\delta\CO$.
We will also make use of the previous determination of $r_m$.

We first consider the axial rotation properties of the
flavor-singlet operators. The singlet axial current
should be invariant under non-singlet axial transformations, 
so we enforce
\begin{equation}
\langle (\delta^{(12)}_{A} \CS)\ \widehat{\tr\;A}_\mu(y)  J^{(21)}(z) \rangle 
= O(a^2) 
\,,\qquad [j\ne k]\,.
\label{eq:AsingletWI}
\end{equation}
We choose flavor indices 1 and 2 so that we can use
the standard Gell-Mann basis of $SU(3)$ matrices---permutations
of indices are, of course, allowed.
Since the right-hand side of
eq.~(\ref{eq:AsingletWI}) vanishes (at the order we are working),
this relation can only determine 
relative normalizations between independent operators
appearing on the left-hand-side. 
Using eqs.~(\ref{eq:trAimp}) and (\ref{eq:singletM}), one finds
\begin{eqnarray}
\widehat{\tr\; A}_\mu &\propto&
\tr A_\mu + a \bar c_A \partial_\mu \tr P + a d_A \tr(M A_\mu) 
\nonumber \\
&\propto&
\tr A_\mu + a \bar c_A \partial_\mu \tr P 
+ a (d_A/6) \left[3 \tr (\lambda_3 M)\tr (\lambda_3 A_\mu)
                        + \tr (\lambda'_8 M)\tr (\lambda'_8 A_\mu) \right]
\,.
\label{eq:trAmu}
\end{eqnarray}
In the second line we have used 
\begin{equation}
\tr(M \CO) = \frac13\tr(M) \tr(\CO) + \frac12\tr(\lambda_3 M)\tr(\lambda_3 \CO)
+ \frac16\tr(\lambda'_8 M)\tr(\lambda'_8 \CO)
\,,
\end{equation}
valid for diagonal mass matrices,
to express the result in terms of independent operators,
and absorbed the contribution proportional to 
$\tr M\tr A_\mu$ into the overall constant.
In the chiral limit only the first two terms in
(\ref{eq:trAmu}) are present
and so $\bar c_A$ can be determined.
Away from the chiral limit, we must avoid contact terms,
which are proportional to $\widetilde m_{12}$.
Following the discussion above, we do so by
setting $\widetilde m_{12}=0$, leading to
$\tr(\lambda'_8 M) = - 2 r_m \tr M$,
so that eq.~(\ref{eq:trAmu}) becomes
\begin{equation}
\widehat{\tr\; A}_{\mu}\bigg|_{\widetilde m_{12}=0} \propto
\tr A_\mu + a \bar c_A \partial_\mu \tr P 
+ a (d_A/6) \left[3 \tr (\lambda_3 M)\tr (\lambda_3 A_\mu)
                        -2 r_m \tr M \tr (\lambda'_8 A_\mu) \right]
\,.
\end{equation}
Thus $d_A$ can be determined by tuning the 
cofficients of either $\tr(\lambda_3 A_\mu)$
or (assuming $r_m$ is known) $\tr(\lambda'_8 A_\mu)$. 
Both require non-degenerate quarks
(degenerate quarks are not sufficient since
the constraint $\tilde m_{12}=0$ then
implies $\tr(\lambda_3 M)=\tr M=0$),
but $2+1$ flavors suffice ($m_1=m_2\ne m_3$
to determine the coefficient of $\tr(\lambda'_8 A_\mu)$,
and $m_1\ne m_2=m_3$ for that of $\tr(\lambda_3 A_\mu)$).
The net result is the first determination of both
$\bar c_A$ and $d_A$. 

An almost identical discussion holds for the flavor singlet vector
bilinear, the conclusion from which is that one can determine
$\bar c_V$ (for the first time) and $d_V$ (which checks the
determination from the vector charge).

\bigskip
For the other three bilinears the Ward identities are different,
since the singlet bilinears are not invariant:
\begin{eqnarray}
\langle (\delta^{(jk)}_{A}\CS)\ \widehat{tr P}(y) J^{(kj)}(z) \rangle &=& 
2 \langle \widehat{S}^{(jk)}(y)J^{(kj)}(z) \rangle + O(a^2) \,,
\label{eq:singletPWI} 
\\
\langle (\delta^{(jk)}_{A} \CS)\  \widehat{\tr T}_{\mu\nu}(y) 
J^{(kj)}(z) \rangle &=& 
-\epsilon_{\mu\nu\rho\sigma}
\langle \widehat{T_{\rho\sigma}}^{(jk)}(y)J^{(kj)}(z) \rangle 
+ O(a^2)\,, 
\label{eq:singletTWI}
\\
\langle (\delta^{(jk)}_{A} \CS)\ \widehat{\tr S}(y) J^{(kj)}(z) \rangle &=& 
2 \langle \widehat{P}^{(jk)}(y)J^{(kj)}(z) \rangle + O(a^2)\,.
\label{eq:singletSWI}
\end{eqnarray}
The general strategy to enforce these relations is
to take the ratio of the two sides, and require that
the result is unity independent of $y$ and $J$.
This should also be true independent of quark masses,
as long as one avoids contact terms by 
keeping $\widetilde m_{jk}=0$.

Consider first the transformation of the singlet pseudoscalar,
(\ref{eq:singletPWI}), and set $j=1$, $k=2$ for convenience.
If we work at $\widetilde m_{12}=0$ to avoid contact terms,
we find
\begin{align}
\widehat{\tr P} = Z_P r_P \big[ 1 &+ a (\bar d_P + d_P/3) \tr M \big]
\bigg\{\tr P + a g_P \Tr(F_{\mu\nu}\widetilde F_{\mu\nu}) \notag \\
&+ a d_P \big[ \tr(\lambda_3 M) \tr(\lambda_3 P)/2 
 - a r_m \tr M \tr(\lambda'_8 P)/3 \big] \bigg\} + O(a^2) \,.
\end{align}
This implies that we can determine $g_P$ (in the chiral limit)
and $d_P$ (in two independent ways as for $d_A$ above, both
requiring only $2+1$ flavor simulation).
Bringing the overall factor to the RHS of (\ref{eq:singletPWI})
we obtain (again, with $\widetilde m_{12}=0$):
\begin{equation}
\frac{\widehat S^{(12)}}
{Z_P r_P \left[1+a \tr M (\bar d_P+d_P/3)\right]}
=
\frac{Z_S}{Z_P r_P}
\left[1 + a\tr M (\bar b_S - \bar d_P - (r_m-1) b_S/3 - d_P/3)
\right] S^{(12)}
\,.
\label{eq:flavorsingletPRHS}
\end{equation}
Thus we can determine $Z_S/(Z_P r_P)$ (which serves as a check)
and the combination multiplying $\tr M$ (the latter again
requiring only $2+1$ flavors).

The Ward identity just discussed gives the first determination of
$d_P$. This then allows the linear combinations of constants for 
the masses and pseudoscalar bilinears determined previously
(and given in the second section of Table~\ref{tab:summI}) 
to be simplified.
In particular, we can now separately determine
$b_P$, $f_P$, $b_m$ and $d_m$ as well as $d_P$.
This leaves three combinations that cannot yet
be disentangled: $\bar b_m + \bar d_P$,
$\bar b_P- \bar d_P$, and $\bar b_m-\bar d_m$.
Note that, at this stage, combination multiplying $\tr M$ in
eq.~(\ref{eq:flavorsingletPRHS}) does not determine any
further coefficients since $\bar b_S - \bar d_P= -(\bar b_m + \bar d_P)$.
 
The same analysis goes through for the tensor bilinear,
leading to the first determination of $r_T$, $\bar c_T$, $d_T$
and $\bar b_T - \bar d_T - (r_m-1) b_T/3$. Since we know
$b_T$ from previous Ward identities, 
we can extract $\bar b_T-\bar d_T$.

The analysis for the scalar bilinear is more subtle,
due to the presence of the identity operator in $\widehat{\tr S}$.
The net result, however, is as if this identity component was absent,
and we find that one can determine the same list as for the pseudoscalar
after permuting $P\leftrightarrow S$: $g_S$, $d_S$,
$Z_P/(Z_S r_S)$ and (using the previously determined $b_S$)
$\bar b_P-\bar d_S$. Of these, only $d_S$ is new.\footnote{%
Using the relations in Table~\ref{tab:relnI},
$\bar b_P - \bar d_S = (\bar b_P-\bar d_P) + (\bar d_P+\bar b_m) 
+ 2 (\bar d_m -\bar b_m) + 2 d_m/3-2b_m/3$, and all the coefficients
appearing on the RHS are known.}
Using the relation $d_S=b_S + 3\bar b_S$ from Table~\ref{tab:relnI},
and the previously determined $b_S=-2b_m$, we can extract 
$\bar b_S=-\bar b_m$. This allows one to disentangle the remaining
linear combinations of $M$ and $P$ improvement coefficients,
so that we can determine $\bar b_m$, $\bar d_m$, $\bar b_P$ and $\bar d_P$
separately.

We now return to the identity operator contribution
on the LHS of eq.~(\ref{eq:singletSWI}).
Formally, this contribution vanishes since it
it is invariant under axial transformations.
However, this fails on the lattice because the
variation of the identity operator
does not vanish fast enough to overcome
the $1/a^3$ divergence in its coefficient:
\begin{equation}
\langle (\delta^{(jk)}_{A}\CS)\         J^{(kj)}(z) \rangle = O(a^2)
\,.
\end{equation}
To overcome this one can explicitly subtract
the disconnected contribution
\begin{equation}
\langle \widehat{\tr S}(y)\rangle \times 
\langle (\delta^{(jk)}_{A}\CS)\ J^{(kj)}(z) \rangle 
\end{equation}
from the LHS of eq.~(\ref{eq:singletSWI}).
This is equivalent to enforcing the difference of two continuum
Ward identities, with coefficients chosen so as to completely remove
the contribution from the $e_S$ term,
including its mass dependence. The cancellation between
the two terms is between contributions proportional
to $1/a^3$ leaving a residue that must be accurate
to $O(a^2)$.\footnote{%
Note that all three operators in $\widehat{\tr S}$, i.e. 
$\tr S$, $\Tr(F_{\mu\nu} F_{\mu\nu})$,
and $\tr(M S)$, lead to contributions to
$\langle\widehat{\tr S}(y)\rangle$ which
are separately divergent, but each can
be combined with its subtraction term and then multiplied
by the unknown coefficient to be determined.
}
Thus it will require good statistical control.
On the other hand, the dominant contribution to the two
terms will be correlated (since it involves the identity operator),
which will help the cancellation.\footnote{%
This situation is similar 
to the subtraction of power divergent mixing in weak matrix
element calculations {\em in the quenched approximation}, 
which also benefits from correlations
between the two quantities being subtracted, and has been
successfully carried out in 
practice~\cite{kilcuppekurovsky,CPPACSepp,RBCepp}.}
%

In addition to allowing the separation of all the coefficients
for $M$ and $P$, this Ward identity gives a new method for calculating
$b_g$, using the relation $b_g=2 g_0^2 g_S$ and the determination
of $g_S$. Note that this determination can be carried out
in the chiral limit, and so one does not need to 
know $b_g$ {\em a priori}. This makes the determination
of improvement coefficients using the Ward identities 
discussed here self contained.

\bigskip

The final Ward identity giving new information is
\begin{equation}
\langle (\delta^{(jk)}_{A}\CS)\ \widehat T_{\mu\nu}^{(kj)}(y)\ 
                                J(z) \rangle = 
 -\half\epsilon_{\mu\nu\rho\sigma} 
\left\langle \left[
\widehat T_{\rho\sigma}^{(jj)}(y)+\widehat T_{\rho\sigma}^{(kk)}(y)
\right] J(z) \right\rangle + O(a^2)
\label{eq:diagTWI}
\end{equation}
In particular, the RHS contains the diagonal non-singlet,
and thus provides access to $f_T$ for the first time.
To see what we learn from enforcing this identity,
we divide through by
\begin{equation}
Z_T \left[1 + a\,\tr M \bar b_T + a m_{jk} b_T \right]
\xrightarrow[\widetilde m_{jk}=0]{}
Z_T \left[1 + a\,\tr M \left(\bar b_T - \frac{r_m-1}{3} b_T\right) 
\right] \,.
\end{equation}
Then we know the quantities on the LHS of the Ward identity.
The operator on the RHS becomes (again setting $j=1$ and $k=2$,
and still working at $\tilde m_{12}=0$)
\begin{align}
\frac{\widehat T_{\rho\sigma}^{(11)} + \widehat T_{\rho\sigma}^{(22)}}
{Z_T \left[1 \!+\! a \tr M \left(
\bar b_T \!-\! \frac{r_m\!-\!1}{3} b_T\right) \right]}
= \bigg(\frac{2 r_T}{3} &+ \CN a\,\tr M\bigg)(\tr T_{\rho\sigma})^I
+ \left(\frac13 + \CP\,a\,\tr M \right) 
  \tr(\lambda'_8 T_{\rho\sigma})^I \notag \\
&+ \CQ\,a\,\tr (\lambda_3 M)\tr(\lambda_3 T_{\rho\sigma}) \,, 
\end{align}
where
\begin{align}
\CN &= \frac23\left[
r_T \left(\bar d_T - \bar b_T + \frac{d_T}{3} + \frac{r_m-1}{3} b_T\right)
- r_m \left(f_T + \frac{b_T}{3}\right) \right]\,,\\
\CP &= \frac29 r_m (b_T - r_T d_T) \,,\\
\CQ &= \frac16 (b_T + 2 r_T d_T) \,.
\end{align}
Thus in the chiral limit we can determine $r_T$, $\bar c_T$
and $c_T$, which provide cross-checks, while away from the chiral limit
we obtain the three combinations $\CN$, $\CP$ and $\CQ$ (all
obtainable with $2+1$ flavors).
These in turn can be combined to give $b_T$ and $d_T$ separately,
as well as the quantity $r_T(\bar d_T-\bar b_T) - r_m f_T$.
Given the determination of $\bar d_T-\bar b_T$ from the Ward identity
(\ref{eq:singletTWI}) above, we can extract $f_T$.
We cannot, however, see any way of disentangling $\bar d_T$ and $\bar b_T$.

Although we have avoided contact terms by working at $\widetilde m_{12}=0$,
we note that these terms are more complicated here than in the previous 
Ward identity.
The contact terms arise from the fact that, even if one has
on-shell improved the operator
$\widehat P^{(jk)}$ appearing in $(\delta_A^{(jk)}\CS)$ and
the tensor bilinear $\widehat T_{\mu\nu}^{(kj)}$, the product
of these operators at the same position will not be improved.
To improve this product one needs to add all operators with the
same transformation properties as the product, and having appropriate
dimension (here dimension 4 because of the overall factor
of $\widehat m_j+\widehat m_k$). In the Ward identity considered above,
in which the bilinear on the RHS was flavor off-diagonal, 
only a single operator could appear, namely $\delta\CO$ with
appropriate flavor indices. By contrast, in the present identity,
two operators are allowed, namely those appearing on the RHS, 
\begin{equation}
a \widetilde m_{jk} \left(
\delta T_{\mu\nu}^{(jj)}(y) + \delta T_{\mu\nu}^{(kk)}(y)\right) 
\qquad \mathrm{and} \qquad
a \widetilde m_{jk} \tr(\delta T_{\mu\nu})(y) \,.
\label{eq:newEoMT}
\end{equation}
The latter arises from Wick contractions in which the quark and
antiquark in $P^{(jk)}$ are both contracted with the corresponding
antiquark and quark in $T^{(kj)}$. To understand (\ref{eq:newEoMT}) in
terms of operators vanishing by the equations of motion requires a
generalization of  the prescription given in Ref.~\cite{WIPLB}. The
appearance of a second operator plays an important role in the
discussion of vector Ward identities in appendix~\ref{app:VWI}. 

\bigskip
The remaining Ward identities do not provide any new information on the
improvement coefficients, but do provide several important cross-checks.
Consider first
\begin{equation}
\left\langle (\delta^{(jk)}_{A}\CS)\ \widehat{S}^{(kj)}(y)\ 
                                J(z) \right\rangle =  
\left\langle \left[
\widehat{P}^{(jj)}(y)+\widehat{P}^{(kk)}(y)
\right] J(z) \right\rangle + O(a^2) \,.
\label{eq:diagPWI}
\end{equation}
Dividing both sides by
\begin{equation}
Z_S\left[1 + a \tr M \left(\bar b_S - \frac{r_m-1}{3} b_S\right)\right]
\end{equation}
the LHS is then a known quantity, while the RHS becomes
(picking $j=1$, $k=2$ and setting $\widetilde m_{12}=0$)
\begin{align}
\frac{\widehat{P}^{(11)} + \widehat{P}^{(22)}}
{Z_S \left[1 \!+\! a \tr M \left(
\bar b_S \!-\! \frac{r_m\!-\!1}{3} b_S\right) \right]}
= \frac{Z_P}{Z_S}\bigg\{\bigg(\frac{2 r_P}{3} &+ \CN' a\,\tr M\bigg)(\tr P)^I
+ \left(\frac13 + \CP' a\,\tr M \right) \tr(\lambda'_8 P)^I \notag \\
&+ \CQ' a\,\tr (\lambda_3 M)\tr(\lambda_3 P) \bigg\} \,,
\end{align}
where
\begin{align}
\CN' &= \frac23\left[
r_P \left(\bar d_P - \bar b_S + \frac{d_P}{3} + \frac{r_m-1}{3} b_S\right)
- r_m \left(f_P + \frac{b_P}{3}\right) \right] \,,\\
\CP' &= \frac19 \left[ 3(\bar b_P-\bar b_S) + b_P-b_S 
        + r_m (b_P + b_S - 2r_P d_P) \right] \,,\\ 
\CQ' &= \frac16 (b_P + 2 r_P d_P) \,.
\end{align}
A similar analysis holds for the Ward identity
 with $S$ and $P$ interchanged,
except that one must subtract the disconnected component
from both sides of the equation. Combining these two Ward
identities allows the determination of the coefficients listed
in Table~\ref{tab:summI}.

In fact, the disconnected component on the LHS of the Ward
identity (\ref{eq:diagPWI}) with
$S$ and $P$ interchanged gives access to the quark condensate:
\begin{equation}
\frac{1}{Z_P}\left\langle (\delta^{(jk)}_{A}\CS)\ \widehat{P}^{(kj)}(y)
                         \right\rangle \equiv
\frac{1}{Z_P}\left\langle 
\widehat{S}^{(jj)}(y)+\widehat{S}^{(kk)}(y)
\right\rangle + O(a^2) \qquad \widetilde m_{jk}=0\,.
\label{eq:cond}
\end{equation}
Here the idea is to calculate the LHS, and use it as the
definition of the RHS. We know all the improvement and renormalization
constants appearing on the LHS, and, by working at $\widetilde m_{jk}=0$
we avoid contact terms. In this way we can calculate the
improved condensate, including its dependence on $\hat m_l$.
This is the $O(a)$ improved version of
the method first suggested in Ref.~\cite{Bochicchio}.

The final two Ward identities of this type involve vector and axial currents,
e.g.
\begin{equation}
\left\langle (\delta^{(jk)}_{A}\CS)\ \widehat{A}_\mu^{(kj)}(y)\ 
                                J(z) \right\rangle =  
\left\langle \left[
\widehat{V}_\mu^{(kk)}(y)-\widehat{V}_\mu^{(jj)}(y)
\right] J(z) \right\rangle + O(a^2) \,.
\label{eq:diagVWI}
\end{equation}
Now only flavor non-singlet operators appear, so we cannot obtain
information about singlet improvement coefficients.
In fact, this identity and that with $V\leftrightarrow A$
only differ from the previous non-singlet axial Ward identities,
eq.~(\ref{eq:AT}),
when $m_j \ne m_k$. In this case there are additional
quark disconnected contractions absent in (\ref{eq:AT}).
Since we are assuming that we know the improved off-diagonal
axial current, the LHS of this relation is completely known,
allowing a complete determination of the constants appearing 
on the RHS.
It is a straightforward exercise using the form
of the improved diagonal non-singlet bilinears,
eq.~(\ref{eq:diagonalO}),
to show that the new information that one obtains here
over that obtained using eq.~(\ref{eq:AT}),
is a direct determination of $f_V$ and $b_V$.
In particular, this Ward identity provides the only cross check
of the calculation of $f_V$.
The identity with $V\leftrightarrow A$ similarly provides
an alternative determination of $f_A$.

\section{Summary and conclusions}
\label{sec:conc}

Improvement in the presence of non-degenerate dynamical quarks
is far more complicated than that with degenerate quarks
or that in the quenched approximation. Nevertheless, the
considerable number of extra improvement coefficients that
arise for quark bilinears and quark masses
can almost all be determined by enforcing the
vector and axial transformation properties of these operators.
The results from the Ward identities considered above
are collected in Tables~\ref{tab:relnI} and \ref{tab:summI}.
The only scale independent quantities which are left undetermined
are $\bar d_A$ and 
$\bar b_T+\bar d_T$.\footnote{%
This corrects the conclusion of Ref.~\cite{lat99} that there
were three undetermined scale independent quantities.}
To determine these one must use other methods.
The only two that we are aware of are improved non-perturbative
renormalization~\cite{NPR,CNGI}, and matching to perturbative
forms for short distance correlation functions of 
bilinears~\cite{Martinelli}.
The scale-dependent quantitites 
$Z_T$, $Z_S Z_P$, and $r_A$ are also undetermined,
but this had to be the case as
Ward identities do not involve a renormalization scale.
For these one must use a method like non-perturbative 
renormalization.

It is interesting to understand 
why the two scale independent quantities
cannot be determined using Ward identities.
In the case of $\bar d_A$, the reason is the lack of an identity involving
$\tr A_\mu$ that has a non-vanishing variation.
Thus the overall normalization factor, which includes $\bar d_A$,
cannot be determined. This is clearly related to the fact that
$r_A$ cannot be determined, because it is scale-dependent.

The reason is similar for $\bar b_T+\bar d_T$.
Ward identities
relate the flavor singlet tensor to the flavor non-singlet tensor.
Overall factors thus cancel, and the mass dependent
part of the overall factor
is proportional to $\bar b_T+\bar d_T$.
Thus, in essence, this combination cannot be determined
because $Z_T$ cannot.

To test these arguments, and for completeness,
we have extended the analysis to two and four 
non-degenerate flavors. 
These cases are also of phenomenological interest. 
A summary of the results is given in appendices~\ref{app:fourflavors}
and \ref{app:twoflavors}. 
With four flavors, one might naively have hoped to determine
more coefficients, since the three flavor theory is 
included as a subset.
We find, however, that although most Ward identities by themselves allow
more combinations of coefficients to be determined,
so that the analysis is cleaner,
 the final result is the same as with three flavors.
In particular, $\bar d_A$ and $\bar b_T+\bar d_T$ cannot
be determined, for exactly the same reasons as for three flavors.

With two flavors, there are,
on the one hand, fewer coefficients to determine,
but, on the other hand, fewer Ward identities.
Furthermore, each identity
determines fewer coefficients because there are less masses to
vary independently. The net result is that there are more 
undetermined combinations of scale independent quantities
than for three or four flavors (eight in all).
Particularly striking is the result that
 one cannot determine the improved 
{\em flavor non-singlet} axial current, pseudoscalar density or tensor
bilinear away from the chiral limit.

Given the complexity of the calculations we outline, one might
wonder about possible simplifications. Since adding flavors
extends ($N_f=2\rightarrow 3$) or simplifies ($N_f=3\rightarrow 4$)
the analysis, one might consider using a partially quenched simulation
with, say $2+1$ flavors of sea quarks, and four or more flavors of
valence quarks. To add additional information over the unquenched
analysis, one must necessarily consider theories with differing sea
and valence content, which are therefore not unitary. This makes the
basis of the Symanzik improvement program less secure. Nevertheless,
it is certainly possible, assuming the improvement program remains valid,
to use the enlarged graded symmetry groups of
partially quenched theories to constrain
the allowed improvement coefficients, and to generalize the analysis presented
here.

Staying within the unquenched three flavor theory, one can ask whether
it is sufficient to use $2+1$ flavor theories in which at least two quarks are degenerate.
This point has been discussed for each identity considered in the text,
and we find that $2+1$ flavors is sufficient to determine all the
allowed coefficients.

Another practical question is whether one needs to determine all the
improvement coefficients for phenomenologically interesting applications.
Particularly interesting are (1) matrix elements of the electromagnetic
current, (2) matrix elements of flavor off-diagonal
vector and axial currents (for weak transitions),
(3) quark masses and (4) matrix element of the mass term in the
action, $m_j S^{(jj)}$. We consider these cases in turn.
(1) The improvement and renormalization
coefficients for the electromagnetic current,
which is a flavor non-singlet for $N_f=3$, are $Z_V$,
$b_V$, $\bar b_V$ and $f_V$,
and these can all be determined by normalizing the matrix elements of the
charge. The determination does, however, require quark-disconnected
 matrix elements. This also determines the improved flavor off-diagonal
vector current.
(2) The off-diaogonal axial current requires $Z_A$, $b_A$ and $\bar b_A$,
which can all be determined from flavor off-diagonal three point
Ward identities, as long as we know $b_V$. Thus only quark-connected 
contractions are needed.
(3) Improvement of individual quark masses requires $Z_m$, $r_m$,
$b_m$, $\bar b_m$, $d_m$ and $\bar d_m$, i.e. both flavor singlet and
non-singlet coefficients [see eq.~(\ref{eq:singleM})]. 
To determine these requires all the types of
Ward identity we consider, i.e. two and three-point Ward identities involving
both flavor singlet and non-singlet operators. The same is true for
(4), the combination $m_j S^{(jj)}$. 
Thus we see that for some applications there are considerable simplifications,
but for others there are not.
We stress however, that for all of these quantities one must first
determine $b_g$, although, as discussed in the introduction, a perturbative
determination of this numerically small coefficient may be sufficient.

Finally, we note that one might consider our work as
an advertisement for other
approaches to improvement with Wilson-like fermions,
namely ``Wilson averaging''~\cite{FrezzottiRossi}
and twisted mass QCD\cite{tmqcd} at maximal twist~\cite{FrezzottiRossi}.
In both approaches the $O(a)$ terms are {\em automatically} absent
in the physical matrix elements of the operators we consider here,
so that no improvement of the operators themselves is necessary.
For this to hold, however, one needs an even number of fermions,
and so for the physical case one must simulate with four flavors.
First work in this direction has begun~\cite{fourflavortwisted}.

\section*{Acknowledgments}
\label{sec:acknowledge}
We thank Rainer Sommer for comments on the manuscript.
This work was supported in part by US Department
of Energy through grants
DE-FG03-96ER40956/A006
and
KA-04-01010-E161,
and by the BK21 program at Seoul National
University, the SNU foundation \& Overhead Research fund and the
Korea Research Foundation through grant KRF-2002-003-C00033.

\appendix

\section{Four non-degenerate flavors}
\label{app:fourflavors}

In this section we briefly describe the generalization 
of our calculations to $N_f=4$. Having one additional
flavor does not change the number of improvement coefficients,
but does increase the number of independent masses that
one can vary. This allows the extraction of
additional improvement coefficients in many of the
Ward identities.
The final result, however, is the same as for $N_f=3$: all
scale independent coefficients can be determined except
for $\bar d_A$ and $\bar b_T + \bar d_T$.

We summarize the results in Tables~\ref{tab:relnI}
and \ref{tab:fourflavors},
and in the following comment on the ways in which the
calculation differs from that with $N_f=3$.

\begin{table}[htbp]
\caption{Normalization and improvement coefficients determined
using various Ward identities (which are denoted schematically)
for $N_f=4$. Notation as in Table~\protect\ref{tab:summI}.}
\label{tab:fourflavors}
\begin{center}
\begin{tabular}{ p{2in} p{1.4in} p{2.8in} }
Ward identity & LO& NLO
\\ \hline
$\langle H | \sum_{\vec x} \hat V_4^{(jj)} | H\rangle = Q_H^j$ 
&
$Z_V$, $r_V$ 
& 
$b_V$, $\bar b_V$, $f_V$, $d_V$, $\bar d_V$ 
\\ \hline
$\partial_\mu \hat A_\mu^{(jk)} = (\hat m_j + \hat m_k) \hat P^{(jk)}$ 
&
$Z_m Z_P/Z_A$,\ $r_m$ \newline
[$c_{SW}$,\ $c_A$]
&
$b_m$,\ $d_m$,\ $\bar b_m - \bar d_m$, \newline
$b_P - b_A$, $\bar b_P - \bar b_A + \bar d_m$
\\ \hline
$\partial_\mu (\hat A_\mu^{(jj)} - \hat A_\mu^{(kk)})$ \newline
$=2 \hat m_j \hat P^{(jj)} - 2\hat m_k \hat P^{(kk)}$ 
&
$Z_m Z_P/Z_A$,\ $r_m$ \newline
$r_P$,\ $g_P$, [$c_{SW}$,\ $c_A$]
&
$b_A$,\ $f_A$,\ $b_P$,\ $f_P$,\ $d_P$,\ $b_m$,\ $d_m$, \newline
$\bar b_P - \bar d_P$,\ $\bar b_A - \bar d_P - \bar d_m$, 
$\bar b_m - \bar d_m$
\\ \hline
%
%
$\delta_A^{(ij)} V^{(jk)} = A^{(ik)}$ \newline
and $V \leftrightarrow A$ 
& $Z_V$ \newline
  $Z_A^2$, [$c_V$] 
& $b_V$,\ $\bar b_V$ \newline
  $b_A$,\ $\bar b_A$
\\
$\delta_A^{(ij)} T^{(jk)} = T^{(ik)} $
& $Z_A$, [$c_T$] 
& $b_T$,\ $4\bar b_A - b_A (r_m-1)$
\\
$\delta_A^{(ij)} P^{(jk)} = S^{(ik)}$ \newline
and $P \leftrightarrow S$ 
& $Z_S/Z_P$ \newline 
  $Z_A^2$ 
& $b_S$,\ $b_P$,\ $\bar b_P - \bar b_S$ \newline
  $4\bar b_A - b_A(r_m-1)$
\\ \hline\hline
$\delta_A^{(ij)} \tr A = \delta_A^{(ij)} \tr V = 0$ &
$\bar c_A$,\ $\bar c_V$ &
$ d_A$,\ $ d_V$ 
\\
%
%
$\delta_A^{(ij)} \tr P = 2 S^{(ij)}$ &
$r_P Z_P/Z_S$, $g_P$ &
$d_P$,\ 
$4(\bar b_S - \bar d_P) - b_S (r_m-1)$\\
$\delta_A^{(ij)} \tr T = 2 T^{(ij)}$ &
$r_T$, $\bar c_T$ &
$d_T$,\ $4(\bar b_T - \bar d_T) - b_T (r_m-1)$\\
$\delta_A^{(ij)} \tr S = 2 P^{(ij)}$ &
$r_S Z_S/Z_P$, $g_S$ &
$d_S$,\ $4(\bar b_P - \bar d_S) - b_P (r_m-1)$\\
$\delta_A^{(ij)} T^{(ji)} = T^{(ii)} + T^{(jj)}$ &
$r_T$,  $\bar c_T$, [$c_T$]&
$b_T$, $d_T$, $r_T(\bar d_T - \bar b_T) - r_m f_T$\\
$\delta_A^{(ij)} S^{(ji)} = P^{(ii)} + P^{(jj)}$ \newline
and $S \leftrightarrow P$ 
& $Z_P/Z_S$, $r_P$,\ $g_P$ \newline
  $r_S$,\ $g_S$
& $b_P$,\ $b_S$,\ $d_P$,\ $d_S$,\
  $\bar b_P - \bar b_S$ \newline
$r_P(\bar d_P - \bar b_P) - r_m f_P$,\
$r_S(\bar d_S - \bar b_S) - r_m f_S$
\\ 
$\delta_A^{(ij)} A^{(ji)} = V^{(jj)} - V^{(ii)}$  &
Not new &
$\bar b_V$,\ $b_V$,\ $f_V$ \\
$\delta_A^{(ij)} V^{(ji)} = A^{(jj)} - A^{(ii)}$  &
Not new &
$\bar b_A$,\ $b_A$,\ $f_A$ 
\\
\hline
\end{tabular}
\end{center}
\end{table}

\begin{enumerate}
\item
The general forms for the improvement of bilinears
and masses, eqs.~(\ref{eq:nonsingletO},%
\ref{eq:singletO},\ref{eq:nonsingletM},\ref{eq:singletM}),
are unchanged, except that $\lambda$ are now $SU(4)$
generators. This is because no special properties of
$SU(3)$ have been used in writing these equations.
Thus the number of improvement coefficients is unchanged
from $N_f=3$.

\item
The use of the vector Ward identity, eq.~(\ref{eq:vectorWI}),
is also unchanged.

\item
The enforcement of ``$Z_S Z_m = 1$'',
i.e. eq.~(\ref{eq:SMR}), is similar to the
three flavor case, although more complicated
algebraically as it involves determinants of
$4\times 4$ matrices. The results are given
in Table~\ref{tab:relnI}; some are identical to
those with $N_f=3$, while others have $N_f$
dependence. There remains a single constraint,
due to the absence of an ``$f$-term'' in the
improvement of the quark masses.

\item
The two-point axial Ward identities are 
somewhat more powerful than for $N_f=3$, due to
the greater number of independent combinations
of masses that can be constructed, particularly
at quadratic order. 
As can be seen from Table~\ref{tab:fourflavors},
one can determine ten combinations of improvement coefficients,
all of which are fairly simple,
compared to nine when $N_f=3$,
some of which are complicated (see Table~\ref{tab:summI}).
For four flavors the flavor off-diagonal identity, eq.~(\ref{eq:WImassdef}),
becomes redundant, giving  no information not contained in the
flavor diagonal identity (\ref{eq:2ptDAWI}).

\item
Three-point Ward identities involving only 
off-diagonal bilinears,
eq.~(\ref{eq:AT}),  are also more powerful,
because there is one additional combination of
masses that is independent from that multiplying
the contact term.
This allows the separate determination of
the $b_O$ from these identities alone,
and of $\bar b_A$, $\bar b_V$, and $\bar b_S-\bar b_P$.
On the other hand, the net result, including
information from previous identities, is 
the same as for $N_f=3$: the newly
determined constants 
remain $Z_A$, $Z_S/Z_P$, $\bar b_A$ and $b_T$.

\item
The Ward identities involving axial transformation
of flavor singlets, e.g. eq.~(\ref{eq:AsingletWI}),
give essentially the same information as for $N_f=3$.
The extra combination of masses that is available
does not multiply new combinations of coefficients.
These identities continue to provide the only
determination of $d_A$, $\bar c_A$ and $\bar c_V$,
as well as the first determination of $d_S$.
The latter, together with the constraint
$d_S=b_S + N_f \bar b_S$, allows the separate
determination of all the mass and pseudoscalar
improvement coefficients, as for  $N_f=3$.

\item
The final type of identity, 
exemplified by eq.~(\ref{eq:diagTWI}),
is more powerful with $N_f=4$, 
as can be seen from the tables.
Nevertheless, the only new information obtained
is $f_T$.
\end{enumerate}

In summary, with four flavors one has considerably
more cross checks, and the extraction of individual
improvement coefficients is more straightforward,
but in the end one obtains the same set
of coefficients as for three flavors.

\section{Two non-degenerate flavors}
\label{app:twoflavors}

In this appendix we describe how our considerations change when
$N_f=2$. The first new feature is that there is one less independent
improvement coefficient for each bilinear and for the quark
masses. This follows from the fact that 
$d_{abc} \propto \Tr(\sigma_a\{\sigma_b,\sigma_c\}) = 0$ in $SU(2)$
(with $\sigma_a$, $a=1,2,3$ the Pauli matrices). Thus one has  
\begin{align}
\{\sigma_a, M\} = \sigma_a \tr M + \tr(\sigma_a M)\mathbb{1}
\; \Longrightarrow \;\,
&(1)\; \tr\left(\{\sigma_a,M\}\CO\right)
= \tr M \tr(\sigma_a\CO) + \tr(\sigma_a M) \tr\CO \notag \\
&(2)\; \tr(\sigma_a M^2) = \tr(\sigma_a M) \tr M \,.
\end{align}
This means that the $b_O$ term in eq.~(\ref{eq:nonsingletO}),
and the $b_m$ term in eq.~(\ref{eq:nonsingletM}), are not
independent and can be absorbed into the other terms by
changing their coefficients as follows:
\begin{equation}
\bar b_\CO \longrightarrow b'_\CO=\bar b_\CO + b_\CO/2 \,, \qquad
\bar f_\CO \longrightarrow f'_\CO=f_\CO + b_\CO/2 \,, \qquad
\bar b_m \longrightarrow \bar b'_m=\bar b_m + b_m \,.
\label{eq:absorbs}
\end{equation}
In the following, we assume that this has been done, and that the
primes are then dropped.

Although there are thus six less coefficients to determine,
it turns out that there are fewer Ward identities available,
and that each is less powerful than for $N_f=3$.
The results are collected in Tables~\ref{tab:relnI}
and \ref{tab:twoflavors}, and we discuss the salient
features in the following.

\begin{table}[htbp]
\caption{Normalization and improvement coefficients determined
using various Ward identities (which are denoted schematically)
for $N_f=2$. Notation as in Table~\protect\ref{tab:summI}.}
\label{tab:twoflavors}
\begin{center}
\begin{tabular}{ p{2in} p{1.8in} p{2.4in} }
Ward identity & LO& NLO
\\ \hline
$\langle H | \sum_{\vec x} \hat V_4^{(jj)} | H\rangle = Q_H^j$ 
&
$Z_V$, $r_V$ 
& 
$\bar b_V$, $f_V$, $d_V$, $\bar d_V$ 
\\ \hline
$\partial_\mu \hat A_\mu^{(12)} = (\hat m_1 + \hat m_2) \hat P^{(12)}$ 
&
$r_m Z_m Z_P/Z_A$, [$c_{SW}$,\ $c_A$]
&
$d_m$,\ $\bar b_A - \bar b_P - \bar d_m$
\\ \hline
$\partial_\mu (\hat A_\mu^{(11)} - \hat A_\mu^{(22)})$ \newline
$= 2 \hat m_j \hat P^{(11)}\! -\! 2\hat m_k \hat P^{(22)}$ 
&
$r_m Z_m Z_P/Z_A$ \newline 
$r_m/r_P$,\ $g_P$ \newline
[$c_{SW}$,\ $c_A$]
&
$f_A$,\ $r_P d_P + r_m d_m$\newline
$r_m(d_m + 2 \bar d_m + 2 \bar b_P - 2 \bar b_A)$ \newline
$r_P(2 \bar b_m + 2\bar d_P - 2\bar b_A + d_P) + 2 r_m f_P$
\\ \hline
$\delta_A^{(12)} \tr A = \delta_A^{(12)} \tr V = 0$ &
$\bar c_A$,\ $\bar c_V$ &
$ d_A$,\ $ d_V$ 
\\
$\delta_A^{(12)} \tr P = 2 S^{(12)}$ &
$r_P Z_P Z_A/Z_S$, $g_P$ &
$d_P$\\
$\delta_A^{(12)} \tr T = 2 T^{(12)}$ &
$r_T Z_A$, $\bar c_T$, $c_T$ &
$d_T$\\
$\delta_A^{(12)} \tr S = 2 P^{(12)}$ &
$r_S Z_S Z_A/Z_P$, $g_S$ &
$d_S$\\
$\delta_A^{(12)} T^{(21)} = T^{(11)} + T^{(22)}$ &
$Z_A/r_T$,  $\bar c_T$, $c_T$&
 $d_T$\\
$\delta_A^{(12)} S^{(21)} = P^{(11)} + P^{(22)}$ &
$r_P Z_P/(Z_A Z_S)$,\ $g_P$ & $d_P$ \\
$\delta_A^{(12)} P^{(21)} = S^{(11)} + S^{(22)}$ &
$r_S Z_S/(Z_A Z_P)$,\ $g_S$ & $d_S$ \\
$\delta_A^{(12)} A^{(21)} = V^{(22)} - V^{(11)}$  &
$Z_V/Z_A^2$,\ $c_V$,\ [$c_A$] & $f_V$ \\
$\delta_A^{(12)} V^{(21)} = A^{(22)} - A^{(11)}$  &
$Z_V$,\  $c_V$,\ [$c_A$]& $f_A$ 
\\
\hline
\end{tabular}
\end{center}
\end{table}

\begin{enumerate}
\item
The use of the vector Ward identity, eq.~(\ref{eq:vectorWI}),
is unchanged, aside from the fact that there is one less
coefficient to determine.

\item
The enforcement of ``$Z_S Z_m = 1$'',
i.e. eq.~(\ref{eq:SMR}), follows the same
steps as above, but leads to simpler
relations because of the absence of $b_m$ and
$b_S$. The results in Table~\ref{tab:relnI} remain
valid for $N_f=2$ as long as one sets $b_m=b_S=0$.\footnote{%
Alternatively, one can keep the relations as they stand in
the Table and then absorb $b_m$ and $b_S$ into the other
coefficients as in eq.~(\protect\ref{eq:absorbs}).}
\item
The two-point axial Ward identities are 
less powerful than for $N_f=3$, due to
the smaller number of independent combinations
of masses. Combining the flavor diagonal and off-diagonal
identities, one only determines
$r_m Z_m Z_P/Z_A$, $r_P/r_m$ and $g_P$ at LO,
and $f_A$, $d_m$, $d_P$, and the combinations
$\bar d_m + \bar b_P - \bar b_A$ and
$r_P(\bar b_m - \bar b_A + \bar d_P) + r_m f_P$
at NLO. Note that one cannot determine $r_m$ and
$r_P$ separately, nor obtain $b_A$, unlike for
$N_f=3$.
\item
For two flavors there are no three point Ward identities involving only 
off-diagonal bilinears. Thus one loses what has
been one of the central tools in quenched studies.
In particular, for $N_f=3$ and $4$ these are the identities
that are used to determine $c_V$ and $c_T$.
Here we need to use other identities for this purpose,
as discussed below and indicated in the Table.
\item
The Ward identities involving axial transformations
of the singlet axial and vector currents, 
e.g. eq.~(\ref{eq:AsingletWI}),
give the same information as for $N_f=3$.
However, those involving the tensor, scalar and pseudoscalar
bilinears give less information.
This is because there is one less combination of
quark masses that is independent of the contact term.
In fact the analysis is more straightforward,
because the contact term is proportional to
$\widetilde m_{12} \propto \tr(M) + O(a)$, and so
one does not need to know $r_m$ in order to work
at $\widetilde m_{12}=0$.

At this stage we still have not determined
$Z_A$ or $\bar b_A$, so we do not know the normalization
of the axial variation of the action.
Note also that the tensor Ward identity allows the
first determination of $c_T$.
\item
The final type of identity is that
exemplified by eq.~(\ref{eq:diagTWI}).
This again is simpler to analyze for $N_f=2$, since the
RHS is a pure flavor singlet (for $T$, $S$ and $P$),
rather than a mix of singlet and non-singlet.
In particular, new information is obtained in the chiral limit,
and allows one, combined with previous results,
to disentangle $Z_A$ and $r_T$.

The identities involving the vector bilinears
are the first to allow a determination of $c_V$.
In fact, in the chiral limit, where one works
to determine $c_V$, these identities involve
the same quark contractions as the three-point Ward
identities with only off-diagonal bilinears that
are present for $N_f\ge 3$. Thus, from a computational
point of view, there is no difference in the method
to be used to determine $c_V$ for $N_f=2$.  
This is not, however, the case for $c_T$.
\end{enumerate}

The following scale independent constants
remain undetermined by the Ward identities:
$\bar b_A$ and $\bar d_A$; $\bar b_T$, $\bar d_T$
and $f_T$; $Z_S/Z_P$, $r_P$, $\bar b_P$,
$\bar d_P$ and $f_P$; and $r_S$ and $\bar d_S$ or equivalently
$r_m$ and $\bar d_m$, although the following combinations
of these coefficients 
are known:
\begin{equation}
\bar b_A - \bar b_P - \bar d_m\,,\quad
r_P (\bar b_A - \bar b_m - \bar d_P) - r_m f_P\,,\quad
r_P Z_P/Z_S\,,\quad
r_P r_S \,.
\end{equation}
Thus in total there are eight undetermined combinations
of scale independent coefficients. This is six more
than for $N_f=3$,
despite the need to determine six fewer coefficients.

What is perhaps most striking about this list is that,
even if one uses non-degenerate quarks when implementing
Ward identities, one cannot determine all the coefficients
needed for flavor non-singlet bilinears composed of degenerate
quarks. In particular, since one cannot separately 
determine $\bar b_A$, $\bar b_P$ or $\bar b_T$,
one does not know the overall normalization of the corresponding
bilinears away from the chiral limit.
To determine this one must use a method such as those
proposed in Ref.~\cite{Martinelli,CNGI}.

\section{Vector Ward identities}
\label{app:VWI}

In this appendix we explain the claim
made in the text that, 
other than the normalization of charges,
enforcing vector Ward identities leads to no
information on improvement coefficients. 
It is interesting to see how this works in detail,
and how we concluded otherwise in Ref.~\cite{lat99}.
It also gives a good example of how the mnemonic
introduced in Ref.~\cite{Martinelli,WIPLB}
of using equations-of-motion operators to
determine the form of contact terms
needs to be modified.

We begin by recalling some results from the main text.
Enforcing the charge of hadrons, eq.~(\ref{eq:vectorWI}),
determines the improved, renormalized diagonal
vector currents, $\widehat V_\mu^{(jj)}$,
aside from the $c_V$ and $\bar c_V$ terms. The latter
terms do not contribute to the divergence, so 
$\partial_\mu \widehat V_\mu^{(jj)}$ is fully improved.
This implies that the divergence of the
off-diagonal currents,
$\partial_\mu \widehat V_\mu^{(jk)}$, $j\ne k$,
are also improved, because these involve the same
improvement coefficients as the diagonal currents
(in particular, $Z_V$, $r_V$, $b_V$ 
and $\bar b_V$; the coefficient $f_V$ is known but not required).
These results hold whatever the precise form of
the underlying lattice current.

Next we recall the form of the exact lattice vector Ward
identities. Making the change of variables
$\delta_V \psi_j = \psi_k$ and $\delta_V \bar \psi_k = - \bar \psi_j$
over a region of the lattice $\CV$, one finds
\begin{equation}
\left\langle
a^4 \sum_{\CV}
\left[ (m_j - m_k) S^{(jk)} - \partial_\mu V_\mu^{lat,(jk)} \right]
\CO(y) J(z) \right\rangle
= \left\langle
[\delta_V^{lat}\CO](y) J(z) \right\rangle
\,,
\label{eq:latticeVWI}
\end{equation}
Here $y$ is in the region $\CV$, while $z$ is not, and 
$\delta_V^{lat}\CO$ is the variation of the operator $\CO$
under the vector transformation.
Note that the bare lattice quark masses appear in (\ref{eq:latticeVWI})
irrespective of the presence of the $c_{SW}$ term,
and that $S^{(jk)}$ is the local scalar bilinear.
The current $V_\mu^{lat,(jk)}$ is the usual
lattice vector current, also unaffected by the $c_{SW}$ term.
It is associated with a link, and not a site, but its divergence
is associated with a site:
\begin{eqnarray}
V_\mu^{lat,(jk)}(x+\frac{\hat\mu}{2}) &\equiv&
\frac12 \left[\bar \psi_j(x) \gamma_\mu U_{x,\mu} \psi_k(x+\hat\mu) +
\bar \psi_j(x+\hat\mu) \gamma_\mu U_{x,\mu}^{\dagger} \psi_k(x) \right]
\,,\nonumber\\
\left[\partial_\mu V_\mu^{lat,(jk)}\right](x)
&\equiv&
\sum_\mu \frac{1}{a}\left[ V_\mu^{lat,(jk)}(x+\frac{\hat\mu}{2})
 - V_\mu^{lat,(jk)}(x-\frac{\hat\mu}{2})\right]
\,.
\label{eq:latticeV}
\end{eqnarray}

The results (\ref{eq:latticeVWI}) and (\ref{eq:latticeV})
hold both for $j\ne k$ and $j=k$. In particular, the diagonal
lattice current is conserved, and it generates canonically
normalized vector transformations. This implies
that the charge constructed from the lattice current is
correctly normalized
[as can be derived from eq.~(\ref{eq:latticeVWI})].
In other words, for this current,
$Z_V^{lat}=1=r_V^{lat}$, and the mass dependent
improvement coefficients all vanish.
As is well known, however, the lattice current is not 
improved.
For this one must add $c_V$- and $\bar c_V$-like terms,
as in eq.~(\ref{eq:improvedV}),
except that they must be associated with a link,
 rather than a site. It is possible to do so
in such a way that taking the divergence of the
current [in the form that appears in the Ward identity
(\ref{eq:latticeVWI})]
exactly cancels the $c_V$- and $\bar c_V$-like terms.\footnote{%
The form of the $c_V$ and $\bar c_V$ terms for
$V_\mu^{lat}(x+\hat\mu/2)$ is
$\partial_\nu [T_{\mu\nu}(x+\hat\mu) + T_{\mu\nu}(x)]/2$,
where the derivative is the symmetric difference.}
Because of this the $c_V$- and $\bar c_V$-like terms
neither contribute to the vector Ward identities nor
affect the values of the other improvement and normalization
constants.

In summary, the divergences of the 
flavor-diagonal conserved lattice vector currents
are automatically improved. As noted above, this also
implies that the divergences of the flavor off-diagonal
lattice vector currents are improved. Thus the
$\partial_\mu V_\mu^{lat,(jk)}$ term in the vector
variation of the action in the lattice
Ward identity (\ref{eq:latticeVWI}) is improved.
This means that it can be replaced, up to errors of
$O(a^2)$, with $\partial_\mu \widehat V_\mu^{(jk)}$,
the divergence of the improved {\em local} vector current.
In this replacement we do not have to worry about
contact terms as $\sum_\CV \partial_\mu \widehat V_\mu$
vanishes except at the surface of the region $\CV$,
and by assumption there are no other operators there.

The other term in the vector variation of the
lattice action in (\ref{eq:latticeVWI})
involves bare masses and scalar densities.
As shown by eq.~(\ref{eq:mS}), this maintains the
same form when written in terms of improved masses and
scalar densities.

Combining these observations we can rewrite the lattice vector
Ward identity in terms of the improved bilinears and masses
considered in the main text:
\begin{equation}
\left\langle
a^4 \sum_{\CV}
\left[ (\widehat m_j - \widehat m_k) \widehat S^{(jk)} 
- \partial_\mu \widehat V_\mu^{(jk)} \right]
\CO(y) J(z) \right\rangle
= \left\langle
[\delta_V^{lat}\CO](y) J(z) \right\rangle + O(a^2)
\,,
\label{eq:improvedVWI}
\end{equation}
This relation leads, amongst other things, to the normalization
condition for the vector charge, eq.~(\ref{eq:vectorWI}).
Note that this relation is ``off-shell" improved, since there are no
contact terms of $O(a)$. These could only enter with the
scalar density term, but the result eq.~(\ref{eq:mS}) is
an algebraic identity, so using it does not
introduce additional terms.
The improved bilinear $\widehat S^{(jk)}$ is, however,
only on-shell improved, and this will be crucial in the
following.

The result eq.~(\ref{eq:improvedVWI}) shows that, if one
uses on-shell improved local bilinears and masses in the 
discretized form of the vector variation of action,
then the operators in the associated Ward identities 
{\em automatically transform with the correct normalization}.
This is the concrete form of the statement made in the text
that the vector Ward identities are automatically satisfied.
There are some subtleties, however, in the application of this
result, and we spend the remainder of this appendix describing
some examples.

\subsection{Two-point vector Ward identities}

In quenched studies of improvement,  the two-point vector
Ward identity 
\begin{equation}
\left\langle
\left[ (\widehat m_j - \widehat m_k) \widehat S^{(jk)}(x) 
- \partial_\mu \widehat V_\mu^{(jk)}(x) \right] J^{(kj)}(0) \right\rangle
=
O(a^2) \,,\qquad j\ne k,\quad x\ne 0
\,,
\end{equation}
has been used as part of the method employed
to determine improvement coefficients~\cite{Petronzio,WIPLB}.
The result eq.~(\ref{eq:improvedVWI}) shows, however,
that this identity is automatically satisfied, as long as
one uses the correctly normalized vector current.
Thus it serves only to check the normalization of the
vector current, and provides no information on the
improvement coefficients of quark masses and scalar bilinears.
This point was not appreciated in Ref.~\cite{WIPLB}.

\subsection{A misleading three-point vector Ward identity}

In Ref.~\cite{lat99}, we argued that we could determine
$g_S$ and $g_P$ by enforcing the vector transformation properties
of the flavor-singlet scalar and pseudoscalar bilinears, respectively.
This is incorrect, as we now show. It is simplest to do this with
the pseudoscalar density as it has no mixing with the identity
operator.

The identity in question is
\begin{equation}
\left\langle
a^4 \sum_{\CV}
\left[ \partial_\mu \widehat V_\mu^{(jk)} \right]
\widehat{\tr P}(y) J^{(kj)}(z)
\right\rangle = O(a^2) 
\,,
\label{eq:deltaVSsinglet}
\end{equation}
where we work in the chiral limit so that there
are no contact terms.
We recall that, in the chiral limit,
the improved, normalized pseudoscalar
density is
\begin{equation}
\widehat{\tr P} = Z_P r_P \left[\tr P + 
a g_P {\mathrm Tr}(F_{\mu\nu}\widetilde F_{\mu\nu})\right]
\,.
\end{equation}
Our previous argument was that, in order for the LHS of
(\ref{eq:deltaVSsinglet}) to be of $O(a^2)$, 
$O(a)$ contributions from the two terms in $\widehat{\tr P}$
must cancel against each other, thus determining $g_P$.
In fact, it follows from eq.~(\ref{eq:improvedVWI}) that
both terms separately are automatically of $O(a^2)$, e.g.
\begin{equation}
\left\langle
a^4 \sum_{\CV}
\left[ 
\partial_\mu \widehat V_\mu^{(jk)} \right]
\tr P(y) J^{(kj)}(z)
\right\rangle = O(a^2) 
\,,
\end{equation}
so that the identity (\ref{eq:deltaVSsinglet}) is
satisfied for any value of $g_P$.
In other words, the two terms in $\widehat{\tr P}$ are separately
invariant under vector transformations up to $O(a^2)$.

\subsection{A paradox and its resolution}
\label{app:paradox}

Another vector Ward identity used in Ref.~\cite{lat99} was
\begin{align}
\left\langle \left[ 
(\widehat m_j - \widehat m_k) \widehat S^{(jk)}(x) 
- \partial_\mu \widehat V_\mu^{(jk)}(x) \right] 
\widehat\CO^{(kj)}(y) J(z) \right\rangle 
&= \left\langle \left[
\widehat\CO^{(kk)}(y)-\widehat\CO^{(jj)}(y)\right]
J(z) \right\rangle \notag \\
&\qquad+ \mathrm{contact\ terms} + O(a^2) \,.
\label{eq:3ptVWI}
\end{align}
Here we work at non-zero quark mass, so there are contact
terms of $O(a)$ because the scalar density is not off-shell improved.
These are proportional to $(m_j-m_k)$ (without any factors involving
$r_m$ as in the axial case), because
$\widehat m_j-\widehat m_k \propto m_j-m_k + O(a)$.
If we use the expressions (\ref{eq:chargedO}) 
and (\ref{eq:diagonalO}) for the improved bilinears,
and divide through by 
\begin{equation}
Z_O [1 +a\bar b_O \tr M + a b_O m_{kj}]
\,,
\end{equation}
the Ward identity we are enforcing becomes
\begin{align}
&\left\langle \left[ 
(\widehat m_j - \widehat m_k) \widehat S^{(jk)}(x) 
- \partial_\mu \widehat V_\mu^{(jk)}(x)\right]
\CO^{(kj),I}(y)J(z) \right\rangle \notag \\
&= \bigg\langle \bigg\{
\left[\CO^{(kk),I}(y)-\CO^{(jj),I}(y)\right]
+ a b_O\frac{(m_k-m_j)}{2}\left[\CO^{(kk)}(y)+\CO^{(jj)}(y)\right]
\notag \\
&\qquad + a f_O (m_k-m_j)\tr O(y) \bigg\} J(z) 
\bigg\rangle + \mathrm{contact\ terms} + O(a^2) \,.
\label{eq:new3ptVWI}
\end{align}
Previously, we argued that the form of the contact term
could be determined by off-shell improving
$S^{(jk)}$ with the addition of a term proportional to the
equations-of-motion operator (\ref{eq:EoMops}),
as in Refs.~\cite{Martinelli,WIPLB}.
This leads to a contact term proportional to
\begin{equation}
\mathrm{contact\ term} \propto
\left\langle a (m_k-m_j)\left[\CO^{(kk)}(y)+\CO^{(jj)}(y)\right] J(z) 
\right\rangle
\,,
\end{equation}
i.e. of the same form as the $b_O$ term.
Thus we concluded that $b_O$ could not be determined
from this Ward identity, but that $f_O$ could, since it
multiplies an independent operator.

To see that this is incorrect, we use the general
result (\ref{eq:improvedVWI}), from which it follows that
\begin{equation}
\left\langle
\left[ (\widehat m_j - \widehat m_k) \widehat S^{(jk)}(x) 
- \partial_\mu \widehat V_\mu^{(jk)}(x) \right] 
\CO^{(kj),I}(y)
J(z) \right\rangle
= 
\left\langle 
\left[\CO^{(kk),I}(y)-\CO^{(jj),I}(y)\right]
J(z) \right\rangle + O(a^2)
\,.
\label{eq:correctVWI}
\end{equation}
Note that the improvement terms (those proportional to $c_O$)
rotate just as the bare operators, so $\CO^I$ rotates as a whole.
Comparing to eq.~(\ref{eq:new3ptVWI}) we see that the first
term on the RHS is obtained automatically, while the $b_O$ and
$f_O$ terms must either cancel against contact terms or vanish.
The discussion above implies that the $b_O$ term is canceled,
but we must have $f_O=0$. This would be a paradoxical
conclusion because we used the vector symmetry in the first place
to conclude that the $f_O$ terms are needed, and yet we have
now used the same symmetry to conclude that they vanish.

In fact, this result is wrong. The flaw in the argument is that
there is an additional contact term in eq.~(\ref{eq:new3ptVWI}) 
proportional to
\begin{equation}
\left\langle a (m_k-m_j)\tr \CO(y) J(z) 
\right\rangle
\,,
\end{equation}
and this is of the right form to cancel with the $f_O$ term.
Thus $f_O$ does not need to vanish, and indeed cannot be determined
from the vector Ward identity, just like $b_O$.
The operator in the new contact term, $\tr\CO$, is allowed because
it appears in the operator product of $S^{(jk)}$ and $\CO^{(kj)}$,
in addition to the other contact term operator
$\CO^{(kk)}+\CO^{(jj)}$. The new operator arises from diagrams in which
both the quark and antiquark in the two operators in the product
are contracted together---the closed quark loop then coupling
to $\tr\CO$ through intermediate gluons.
The presence of this operator shows that the mnemonic of off-shell 
improvement through the addition of equations-of-motion operators
needs to be generalized beyond the considerations of 
Refs.~\cite{Martinelli,WIPLB}. A more straightforward approach is
simply to enumerate the allowed operators using symmetries.
Indeed, one can turn the previous argument around,
and use the fact that symmetry implies the presence of
the $f_O$ terms to imply the existence of the new contact terms.

Finally, we note that these considerations resolve the puzzle
concerning the counting of improvement coefficients
mentioned in the text. The off-diagonal
bilinears require one less improvement coefficient than the diagonal
bilinears. How can this be consistent
with the fact that a vector transformation rotates
 the former into the latter (as in the Ward identities just discussed)?
The answer is provided by the presence of the new contact term, which allows
there to be an $f_O$ term in the diagonal bilinears but not in the
off-diagonal ones.

\end{document}